\documentclass[11pt]{amsart}
\usepackage{amsmath, amsthm, amscd, tikz, amssymb, amsfonts, amsxtra, latexsym}
\usepackage{enumerate}
\usepackage{verbatim}
\usepackage[normalem]{ulem} 
\usepackage{array}
\usepackage[utf8]{inputenc}
\usepackage[T1]{fontenc}
\usepackage{lmodern}
\usepackage{hyperref}
\hypersetup{colorlinks = true,allcolors  = black}
\usepackage{mathtools}
\usepackage{cancel}
\usepackage{xcolor}
\usepackage[a4paper,textwidth=16cm,textheight=23.5cm]{geometry}

\DeclarePairedDelimiter\floor{\lfloor}{\rfloor}
\DeclareMathOperator{\lcm}{lcm}
\def\fq{\mathbb{F}_{q}}
\def\fqs{\mathbb{F}_{q^2}}

\def\cF{\mathcal{F}}

\def\cX{\mathcal{X}}
\def\cC{\mathcal{C}}
\def\cP{\mathcal{P}}

\def\cH{\mathcal{H}}
\def\cL{\mathcal{L}}

\def\N{\mathbb{N}}
\def\Z{\mathbb{Z}}

\newcommand{\hull}{\operatorname{Hull}}
\newcommand{\maxEhull}{\operatorname{maxEhull}}

\newif\ifshowrev
\showrevtrue

\DeclareMathOperator\supp{supp}

\makeatletter
\newcommand{\secondlabel}[1]{\ltx@label{#1}}
\makeatother

\newcommand{\sk}{\smallskip}
\newcommand{\msk}{\medskip}

\theoremstyle{plain}
\newtheorem{theorem}{Theorem}[section]
\newtheorem{lemma}[theorem]{Lemma}
\newtheorem{corollary}[theorem]{Corollary}
\newtheorem{proposition}[theorem]{Proposition}

\theoremstyle{definition}
\newtheorem{definition}[theorem]{Definition}
\newtheorem{example}[theorem]{Example}
\newtheorem{remark}[theorem]{Remark}

\newtheorem{assumption}{Assumption}

\theoremstyle{remark}

\begin{document} 
	\title{Hull variation of algebraic geometry codes under divisor twists}
	\author{Erik Mendoza}
	\address{Instituto de Matemática, Universidade Federal do Rio de Janeiro, Cidade Universitária, CEP 21941-909, Rio de Janeiro, Brazil}
	\email{erik@im.ufrj.br}
	
	\author{Sihem Mesnager}
	\address{Department of Mathematics, University of Paris VIII, F-93526 Saint-Denis; Laboratory Analysis, Geometry and Applications (LAGA), University Sorbonne Cit\'e Paris North, CNRS, UMR 7539, F-93430 Villetaneuse; and T\'el\'ecom Paris, Polytechnic Institute of Paris, 91120 Palaiseau, France}
	\email{smesnager@univ-paris8.fr}
	
	\author{Luciane Quoos}
	\address{Instituto de Matemática, Universidade Federal do Rio de Janeiro, Cidade Universitária, CEP 21941-909, Rio de Janeiro, Brazil}
	\email{luciane@im.ufrj.br}

\begin{abstract}

The \emph{hull} $\hull(\cC)=\cC\cap \cC^{\perp}$ of a linear code
$\cC\subseteq\fq^{n}$, where $\cC^{\perp}$ is its Euclidean dual, is the part of
$\cC$ that the standard inner product cannot see. Its dimension interpolates
between two classical extremes: \emph{linear complementary dual} (LCD) codes,
with $\hull(\cC)=\{0\}$, and \emph{self-dual} codes, with
$\hull(\cC)=\cC=\cC^{\perp}$. The hull is preserved by coordinate permutations, but
its dimension is \emph{not} invariant under scaling
$\cC\mapsto \mathbf{x}\cdot \cC$, $\mathbf{x}\in(\fq^{*})^{n}$. This instability
matters both structurally and algorithmically: via Sendrier's
support-splitting algorithm, it governs the complexity of code equivalence,
and it decides whether a prescribed equivalence class contains LCD codes. We
study the resulting \emph{hull-variation problem}, formalised by Chen, for
algebraic geometry (AG) codes $\cC(D,G)$ over a function field $\cF/\fq$ of
genus $g>0$. There, scaling amounts to twisting $G$ by a principal divisor,
so divisor arithmetic replaces that of vectors in $(\fq^{*})^{n}$. Allowing
twists $G\mapsto G+L$ by arbitrary divisors gives a natural geometric
relaxation of the problem.

Writing the dual of the AG code as $\cC(D,G)^{\perp}=\cC(D,H)$, with $H=D-G+W$
and $W$ a canonical divisor, we see that a twist acts by
$(G,H)\mapsto(G+L,H-L)$. Whenever $\gcd(G+L,H-L)$ is non-special, the hull
dimensions of $\cC(D,G+L)$ and $\cC(D,G)$ differ by exactly
$\tfrac12\big(\deg(|G-H|)-\deg(|G-H+2L|)\big)$. For a genuine scaling of the
code, with $L=-(z)$ a principal divisor, this variation is bounded by
$2[\cF:\fq(z)]$. Under the same non-speciality hypothesis, we also obtain a
new upper bound on the hull dimension, depending only on the divisors $D$ and
$W$, and we characterise exactly when it is
attained by a scaling of it. These bounds shed
new light on Chen's maximal hull dimension for AG codes, and explicit scalings
on Kummer extensions show that the first one is sharp.
We construct explicit twists of $G$ that \emph{raise}, respectively
\emph{lower}, the hull dimension by prescribed amounts, one unit at a time,
and apply them to two classical families. Starting from known LCD codes, we
reach every hull dimension in an explicit range. Starting from a self-dual
code ($q$ even) or a code of maximal hull dimension ($q$ odd), we descend
through every hull dimension down to~$0$, which gives explicit families of
LCD codes of various dimensions.
\end{abstract}

\maketitle


\section{Introduction}\label{sec:intro}
Let $\fq$ denote the finite field with $q$ elements, and let $n \geq 1$.
A \emph{linear code} $\cC$ of length $n$ over $\fq$ is an $\fq$-linear
subspace of $\fq^n$, and three important parameters are associated with a code $\cC$: its length $n$, its dimension $k$ as an $\fq$-vector space, and its minimum (Hamming) distance $d$. We call such a code an $[n, k, d]$-code, or simply an $[n, k]$-code when convenient. The \emph{dual code} of $\cC$ is
\[
  \cC^{\perp} = \left\{ \mathbf{x} \in \fq^n : \langle \mathbf{x}, \mathbf{c} \rangle = 0
  \text{ for all } \mathbf{c} \in \cC \right\},
\]
where $\langle \mathbf{x}, \mathbf{y} \rangle = \sum_{i=1}^{n} x_i y_i$ denotes the
standard (Euclidean) inner product on $\fq^n$. The \emph{hull} of a linear code $\cC\subseteq\fq^n$, introduced by Assmus and Key \cite{AK1990}, is the subspace
$$\hull(\cC)=\cC\cap \cC^{\perp},$$ that is, the part of $\cC$ that is invisible to the Euclidean inner product. Its dimension $\dim_{\fq}\hull(\cC)$ interpolates
between two extreme situations: $\hull(\cC)=\{0\}$ characterises the
\emph{linear complementary dual} (LCD) codes \cite{M1992,CG2015,CMTQP2018}, whereas $\hull(\cC)=\cC$
(respectively $\hull(\cC)=\cC=\cC^{\perp}$) characterises the
\emph{self-orthogonal} (respectively \emph{self-dual}) codes. Since
$\dim_{\fq}\hull(\cC)\le\min\{k, n-k\}\le n/2$ for a code of dimension $k$, the
hull dimension ranges over $\{0,\dots,\lfloor n/2\rfloor\}$, the two endpoints
being precisely the LCD and the self-dual (or, for odd $n$, nearly self-dual)
situations. Beyond this structural role, the hull dimension is a quantity of
genuine algorithmic and cryptographic significance, which we now briefly
recall.

In 1997, Sendrier \cite{S1997} established a foundational link between the
hull dimension and the complexity of equivalence testing: the efficiency of the
support-splitting algorithm \cite{S2000,BOS2019}, which is used to test whether two codes are
equivalent, depends crucially on the dimension of the hull. Roughly speaking,
a small hull makes the algorithm efficient, whereas large hulls, and
self-dual codes in particular, are the hard instances. It is therefore natural
to ask by how much, and in which direction, the hull dimension of a given code
can be moved by passing to an equivalent code.

Given a linear code $\cC \subseteq \fq^n$ and a vector
$\mathbf{x} = (x_1, \dots, x_n) \in (\fq^{*})^n$, we define the code
\[
  \mathbf{x} \cdot \cC = \left\{ (x_1 c_1, \dots, x_n c_n) : (c_1, \dots, c_n) \in \cC \right\},
\]
obtained from $\cC$ by the componentwise product with $\mathbf{x}$.
The codes
$\cC$ and $\mathbf{x} \cdot \cC$ are said to be \emph{equivalent}. Since every $x_i$ is non-zero, they have the
same parameters. In 2023, Chen \cite{C2023} formalised the following question for general
linear codes: given a linear code $\cC \subseteq \fq^n$, how does
$\dim \hull({\bf x} \cdot \cC)$ vary as ${\bf x}$ ranges over vectors defining a
monomial equivalence, and what is
$$\max_{{\bf x} \in (\fq^*)^n} \dim_{\fq}\hull({\bf x} \cdot \cC),$$
the {\it maximal hull dimension} of a code $\cC$. This is now called the
{\it hull-variation problem}, and it has been extended to rank-metric codes
\cite{HJ2025} and to codes over binary extension fields \cite{BDM2026}.

Before specialising this problem, let us observe in passing that a permutation
of the coordinates is an isometry for the standard inner product, so that it
commutes with the passage to the dual and leaves the hull dimension unchanged.
The only part of a monomial equivalence that can modify $\dim\hull$ is
therefore its \emph{diagonal} part, a scaling by a vector
${\bf x}\in(\fq^*)^n$; this is precisely the reason why the hull-variation
problem is formulated in terms of the codes ${\bf x}\cdot \cC$.

In the function field setting, where there is more geometric structure, we
specialise the hull-variation problem. Note that for an algebraic geometry code (AG code for short) $\cC(D,G)$ over
a function field $\cF/\fq$, twisting by a rational function $z$ (i.e. replacing
$G$ by $G-(z)$) gives exactly a monomial equivalence between the codes
$\cC(D,G)$ and $\cC(D,G-(z))$; here $z$ is of course assumed to have neither zeros nor poles at the places in the support of $D$; see
Proposition~\ref{prop2210}\, iii) below. In other words, the scaling vectors
${\bf x}$ of Chen's problem are realised geometrically as evaluation vectors
$(z(P_1),\dots,z(P_n))$ of functions of the function field, and the arithmetic
of divisors takes over the role played by the arbitrary vectors of
$(\fq^*)^n$.

In the context of function fields over finite fields, the hull-variation
problem thus translates into the following question: given a function field
$\cF/\fq$ of genus $g>0$ and an AG code $\cC(D,G)$ over $\cF$, determine the
largest variation of the hull dimension that can be obtained by passing to an
equivalent code, that is,
\begin{equation}\label{problem1}
 \max_{\substack{z\in\cF^*\\ \supp((z))\cap\supp(D)=\emptyset}} |\dim_{\fq} \hull (\cC(D,G+(z)))-\dim_{\fq} \hull (\cC(D,G))|.
 \end{equation}
By Proposition~\ref{prop2210}\, ii), the codes equivalent to $\cC(D,G)$ are
exactly the codes $\cC(D,G')$ with $G'=G+(z)$ for a certain function $z \in \cF^*$ satisfying
$\supp(G')\cap\supp(D)=\emptyset$, so that \eqref{problem1} measures the
amplitude of the hull dimension on the whole equivalence class of $\cC(D,G)$.

Having phrased the problem for twists by principal divisors, it is natural to
relax the constraint and to allow arbitrary twists. More generally, one may
ask how much the hull can be improved by letting the divisor $G$ of the code
$\cC(D,G)$ over $\cF/\fq$ be twisted by a divisor $L$:
\begin{equation}\label{problem2}
\max_{\substack{L \text{ a divisor} \\ \supp(L) \cap \supp(D) = \emptyset}}  |\dim_{\fq} \hull (\cC(D,G+L))-\dim_{\fq} \hull (\cC(D,G))|.
 \end{equation}
Since principal divisors are particular divisors, the quantity
\eqref{problem1} never exceeds \eqref{problem2}. The difference between the
two problems is conceptual: a twist by a principal divisor stays inside the
equivalence class of the code (the length, dimension and minimum distance are
preserved), whereas a twist by a general divisor $L$ leads, in general, to a
code with different parameters. Both maxima are taken over a set of integers
contained in $\{0,\dots,\lfloor n/2\rfloor\}$ and are therefore well defined. Problem
\eqref{problem1} is the one with cryptographic meaning (it stays within the
equivalence class), while \eqref{problem2} is its natural geometric
relaxation, which turns out to be the more tractable of the two and, via
Theorem~\ref{prop_perturbation_hull}, yields information on the first. For twists by principal divisors, we characterise in Theorem~\ref{thm:chenattain} when a scaling attains the largest hull dimension allowed by our upper bound of Proposition~\ref{thm:upper-bound-hull},
and we show on Kummer extensions that the bound of Corollary~\ref{cor:principal}
on the variation under a scaling is sharp (Proposition~\ref{prop:kummerprincipal}).

Much of the existing AG-hull literature focuses on building codes with hull
dimension $0$ (LCD codes, wanted for side-channel-resistant implementations
and in code-equivalence-based cryptography)
\cite{MTQ2017, AMLMRS2024, CLM2024, BDM2024, CMQ2025}. Over the rational function field, that is, a function field of genus zero, the hull has also been investigated in \cite{JYSS2027} and \cite{SEL2026}. For general linear codes, and in particular for MDS codes, codes with hulls of prescribed dimension and their application to entanglement-assisted quantum codes have been studied, among others, in \cite{LCC2019, FFLZ2020, CaLiMe2019}. Recall also that, by \cite{CMTQP2018}, every linear code over $\fq$ with $q>3$ is equivalent to an LCD code; hence the interest of explicit LCD AG codes lies in their parameters and in their explicit algebraic structure, rather than in their mere existence.

In this work, instead of looking
for a divisor $G$ to hit a target hull dimension, we ask how far one can move
an existing $G$ by a principal divisor, or by a general one, to obtain a new
code with a larger or smaller hull (see Theorems~\ref{hull+} and~\ref{hull-}). The point of view is thus dynamical rather
than constructive: the hull dimension is studied as a function on the set of
divisors, and the variations are read off from the arithmetic of the pair
$(G,H)$.

 Our contributions can be summarised as follows.
 
 \begin{enumerate}[i)]
\item \emph{An exact formula.} Under a non-speciality condition on the
greatest common divisor $\gcd(G+L,H-L)$, we compute \emph{exactly} the
variation of the hull dimension when $G$ is replaced by $G+L$; the resulting
bound $\deg|L|$ is attained as soon as $L$ and $G-H$ have disjoint supports
(Theorem~\ref{prop_perturbation_hull}). For a principal twist $L=-(z)$ the
variation is bounded by $2[\cF:\fq(z)]$, see Corollary~\ref{cor:principal}. This
 upper bound can be achieved, see Proposition~\ref{prop:kummerprincipal}.
\item \emph{Prescribed raising and lowering.} In Theorem~\ref{hull+}, we show
how to raise the hull dimension by any prescribed amount up to an explicit
maximum and, in Theorem~\ref{hull-}, how to lower it by any prescribed amount
down to $\deg(E)+1-g$, for a non-special divisor $E\le\gcd(G,H)$.
\item \emph{Explicit families.} These perturbations are made fully explicit on
two classical families. Starting from the LCD codes on Kummer extensions of
\cite{CMQ2025}, we obtain in Theorem~\ref{improvinghull} codes with any hull
dimension in an explicit range. Starting from a self-dual code ($q$ even) or
from a code of maximal hull dimension ($q$ odd) on the Hermitian function
field, we reach every hull dimension down to $0$; see
Theorem~\ref{thm:hermitian} and Examples~\ref{ex:hermitianLCD}
and~\ref{ex:hermitianLCDodd}. Proposition~\ref{prop:kummerprincipal} gives
explicit scalings of the Kummer LCD codes that realise the hull dimensions
$2(q+1)s$ inside a single equivalence class, and Remark~\ref{rem:EAQECC} makes
the resulting entanglement-assisted quantum codes explicit.
\item \emph{Beyond non-speciality.} Without the non-speciality condition on the
gcd, we still obtain in Proposition~\ref{prop:onesided} a one-sided bound on the
decrease of the hull dimension, while Proposition~\ref{prop:general} shows that
the exact formula of i) becomes a lower bound for the variation.
\end{enumerate}

The paper is organised as follows. Section~\ref{sec:preliminaries}
fixes the notation and recalls the necessary background on function fields,
Riemann--Roch spaces, AG codes, equivalence, duality and the hull.
Section~\ref{sec:main} contains the main results under the non-speciality
hypothesis, together with the general raising and lowering constructions.
Section~\ref{sec:kummer} makes these constructions explicit on Kummer extensions and on the Hermitian function field.
Section~\ref{sec:general} treats the case in which the non-speciality of the
gcd is not assumed. Section~\ref{sec:conclusion-perspectives} lists open directions.

\section{Preliminaries: function fields, AG codes and the hull} \label{sec:preliminaries}

In this section we fix notation and recall the background results needed in the sequel. We begin with the language of function fields $\cF/\fq$ and Riemann--Roch spaces
(Sections~\ref{sec:ff}--\ref{sec:rr}), we then recall AG codes, their
equivalence and their duality (Sections~\ref{sec:agc}--\ref{sec:dual}), and we
finish with the two ingredients specific to the hull: the description of
$\cC(D,G)\cap \cC(D,H)$ in terms of $\gcd(G,H)$ and Chen's maximal hull dimension
(Sections~\ref{sec:hullgcd}--\ref{sec:maxhull}). The reader familiar with
Stichtenoth's treatment \cite{S2009} may go directly to Section~\ref{sec:hullgcd},
where the new ingredients begin; the earlier subsections serve mainly to fix
conventions and to isolate the elementary facts about divisors that will be
used repeatedly in Sections~\ref{sec:main} and~\ref{sec:general}. Throughout the paper, we write $\dim_{\fq}\hull(\cC)$ for the hull dimension of a code $\cC$.

\subsection{Algebraic function fields and divisors}\label{sec:ff}
Let $\cF/\fq$ be an algebraic function field in one variable of genus
$g=g(\cF)$. We denote by $\mathcal P_\cF$ the set of places in $\cF$, by
$\Omega_\cF$ the space of Weil differentials in $\cF$, by $v_{P}$ the discrete
valuation of $\cF/\fq$ associated to the place $P\in \cP_\cF$, and by
$\mathrm{Div}(\cF)$ the free abelian group generated by the places in $\cF$.
An element in $\mathrm{Div}(\cF)$ is called a divisor. For a function
$z \in \cF$ we let $(z), (z)_\infty$ and $(z)_0$ stand for the principal,
pole and zero divisors of the function $z$ in $\cF$, respectively. Two
divisors $A,B \in \mathrm{Div}(\cF)$ are equivalent, denoted $A \sim B$, if
they differ by a principal divisor, i.e. $B=A+(z)$ for some $z \in \cF^*$.

We shall repeatedly use the following standard conventions. A place $P$ of $\cF/\fq$ has
degree $\deg (P)=[F_P: \fq]$, where $F_P$ is its residue class field, and it
is \emph{rational} if $\deg (P)=1$. For a divisor $A=\sum_{P}v_P(A)P$ we set
$\deg(A)=\sum_P v_P(A)\deg( P)$ and
$\supp(A)=\{P\in\cP_\cF : v_P(A)\neq 0\}$. 
The set of divisors admits a partial order; given two divisors $A$ and $B$, we write $A\le B$ if
$v_P(A)\le v_P(B)$ for every place $P$, and $A\ge0$ ($A$ is \emph{effective})
if $v_P(A)\geq 0$ for every $P$. A divisor $A$ is \emph{principal} if it is the divisor of a function $z \in \cF^*$, that is, $A=\sum_{P \in \mathcal P_\cF} v_P(z) P$. Principal divisors have degree zero, so
equivalent divisors have the same degree. For every element $z\in\cF\setminus\fq$ we also have
$\deg((z)_0)=\deg((z)_\infty)=[\cF:\fq(z)]$. Finally, the divisor $(\eta)$ of a non-zero Weil differential $\eta$ is 
a \emph{canonical divisor}; all canonical divisors are equivalent to each
other and can be characterised by having degree $2g-2$ and $\ell((\eta))=g$.

The partial order $\le$ on $\mathrm{Div}(\cF)$ turns it into a lattice: any two
divisors admit a greatest lower bound and a least upper bound, computed place
by place. This lattice structure, made explicit in
Section~\ref{sec:hullgcd} through $\gcd$ and $\lcm$, is the algebraic engine
behind the whole paper: the hull of an AG code is governed by the ``common
part'' of the two divisors $G$ and $H$, and perturbing $G$ amounts to moving
inside this lattice.

Having fixed the language of divisors, we now recall the spaces of functions
attached to them, and the notion of speciality which will convert dimensions
of codes into degrees of divisors.

\subsection{Riemann--Roch spaces and the index of speciality}\label{sec:rr}

The Riemann--Roch space of $A\in\mathrm{Div}(\cF)$ is the finite-dimensional
$\fq$-vector space
\[
\cL(A)=\{f\in\cF^{\ast} : (f)\ge -A\}\cup\{0\},
\qquad \ell(A)=\dim_{\fq}\cL(A).
\]
We refer to~\cite{S2009} for the following classical results. If the divisors $A$ and $B$ are equivalent with $B=A+(z)$ for some function $z$ in $\cF^*$, then 
 the Riemann--Roch spaces
$\cL(A)$ and $\cL(B)$ are isomorphic; in particular, $\ell(A)=\ell(B)$.
 For every canonical
divisor $W$, the Riemann--Roch theorem asserts that
\begin{equation}\label{eq:RR}
\ell(A)=\deg (A)+1-g+\ell(W-A),
\end{equation}
and hence $\ell(A)=\deg( A)+1-g$ whenever $\deg (A)\ge 2g-1$. The integer
$i(A):=\ell(W-A)$, which does not depend on the choice of $W$, is the
\emph{index of speciality} of $A$; the divisor $A$ is \emph{non-special} if
$i(A)=0$, and \emph{special} otherwise. We now collect some simple results about non-special divisors.
\begin{lemma}\label{lem:nonspecial}
Let $A,B\in\mathrm{Div}(\cF)$.
\begin{enumerate}[i)]
  \item If $\deg (A)>2g-2$, then $A$ is non-special.
  \item If $A$ is non-special, then $\deg( A)\ge g-1$, with $\ell(A)=0$ when
        equality holds.
  \item If $A\le B$ and $A$ is non-special, then $B$ is non-special.
\end{enumerate}
\end{lemma}
\begin{proof}
 i) If $\deg(A)>2g-2$, then $\deg(W-A)<0$, so $\ell(W-A)=0$.
ii) If $i(A)=0$, then \eqref{eq:RR} gives $0\le\ell(A)=\deg(A)+1-g$, so $\deg(A)\ge g-1$, and $\ell(A)=0$ when $\deg(A)=g-1$.
iii) If $A\le B$, then $W-B\le W-A$, hence $\cL(W-B)\subseteq\cL(W-A)=\{0\}$.
\end{proof}

 The case of a non-special divisor $A$ of degree $g-1$ is of particular relevance here: it accounts for
the vanishing of the hull for the LCD codes of Section~\ref{sec:kummer}. 
Moreover, the converse of Lemma~\ref{lem:nonspecial}\,iii) fails in general: removing a place from a
non-special divisor may yield a special one. This asymmetry between
enlarging and shrinking a divisor is reflected in the contrast between
Theorems~\ref{hull+} and~\ref{hull-}.

We now introduce the algebraic geometry codes, whose parameters are
expressed in terms of dimensions of Riemann--Roch spaces.

\subsection{Algebraic geometry codes}\label{sec:agc}
We start by recalling the definition of an algebraic geometry code $\cC(D,G)$ \cite{G1981} over an algebraic function field $\cF/\fq$. Let
$P_1, \dots, P_n \in \mathcal P_{\cF}$ be pairwise distinct rational places of
$\cF/\fq$, and set $D = \sum_{i=1}^n P_i$. Given a divisor
$G \in \mathrm{Div}(\cF)$ such that $\supp(G) \cap \supp(D) = \emptyset$, the
\textit{algebraic geometry code} (or AG code) $\cC_{\mathcal{L}}(D,G)$ is defined by
\begin{equation}\label{AGcode}\secondlabel{defAGcode}
\cC_{\cL}(D,G)= \{(f(P_1), \dots, f(P_n)) : f \in \cL(G)\}\subseteq \fq^n.
\end{equation}
When no confusion is possible, we simply write $\cC(D,G)$ for $\cC_{\cL}(D,G)$.

The disjointness of $\supp(G)$ and $\supp(D)$ guarantees that every
$f\in\cL(G)$ is regular at each $P_i$ (indeed
$v_{P_i}(f)\ge -v_{P_i}(G)=0$), so that the \emph{evaluation map}
$\mathrm{ev}_D:\cL(G)\to\fq^n$, $f\mapsto(f(P_1),\dots,f(P_n))$, is well
defined. It is $\fq$-linear, its image is $\cC(D,G)$ and its kernel is
$\cL(G-D)$; this explains the dimension formula in \eqref{dimcode} below.
Likewise, a non-zero $f\in\cL(G)$ cannot vanish at more than $\deg (G)$ of the
places $P_i$, which gives the Goppa bound $d\geq n-\deg( G)$.

For the rational function field $\fq(x)$ of genus $g=0$, the choice
$G=(k-1)P_\infty$ gives $\cL(G)=\{$polynomials of degree $\le k-1\}$, and
$\cC(D,G)$ is a Reed--Solomon code with parameters $[n,k,n-k+1]$ whenever
$P_1,\dots,P_n$ are pairwise distinct rational places. AG codes therefore
generalise Reed--Solomon codes to curves of positive genus.

We now recall the basic bounds for the parameters of an AG code. 

\begin{proposition} \cite[Theorem 2.2.2, Corollary 2.2.3]{S2009}
	\label{prop:dimcode}
The AG code $\cC_\cL(D, G)$ over $\fq$ as in \eqref{AGcode} is an $[n, k, d]$-code with
\begin{equation} \label{dimcode}
	k=\ell(G)-\ell(G-D) \qquad \text{ and } \qquad d \geq n-\deg(G).
\end{equation}
Moreover, we have:
\begin{enumerate}[i)]
\item If $\deg (G) < n$, then $k=\ell(G) \geq \deg(G)+1-g$. \msk
\item If $2g-2<\deg(G) < n$, then $k=\ell(G)= \deg(G)+1-g$.
\end{enumerate}
\end{proposition}

\subsection{Scaling equivalence of codes and twists of the divisor}\label{sec:equiv}

Having set up the codes $\cC(D,G)$, we now turn to the transformations of a code
that are relevant for the hull. Let $\cC \subseteq \fq^n$ be a linear code and
${\bf x} =(x_1, \ldots, x_n) \in (\fq^\ast)^n$. The set
	$${\bf x} \cdot \cC :=\{ (x_1 c_1, \ldots, x_n c_n) : (c_1, \ldots, c_n) \in \cC \}$$
is clearly another linear code over $\fq$. Note that the codes $\cC$ and ${\bf x} \cdot \cC$ have the same length, dimension and minimum distance.

\begin{definition}
Two linear codes $\cC_1$ and $\cC_2$ over $\fq$ are \textit{equivalent} if $\cC_2={\bf x} \cdot \cC_1$ for some ${\bf x} \in (\fq^\ast)^n$. In this case we write $\cC_1 \sim_{\bf x} \cC_2$, or simply $\cC_1 \sim \cC_2$ when the vector ${\bf x}$ need not be specified. A linear code $\cC\subseteq\fq^n$ is called \emph{iso-dual} if it is equivalent to its dual, that is, $\cC^\perp={\bf x}\cdot\cC$ for some ${\bf x}\in(\fq^\ast)^n$. 
\end{definition}

In this paper, equivalence is thus meant in the restricted sense of \emph{scalings}: only the diagonal part of a monomial equivalence is retained, which, as observed in the Introduction, is the only part that can affect the hull.

The next proposition translates this algebraic notion into the language of
divisors: scalings of an AG code correspond to twists of $G$ by principal
divisors. We collect well-known results on the equivalence of AG codes; see \cite[Corollaries~2.7 and~3.6]{S1988} and \cite[Propositions~2.2.10 and~2.2.14]{S2009}.

\begin{proposition} \label{prop2210}
Let $\cC(D,G)$ be an AG code with $D=P_1+\cdots+P_n$.
	\begin{enumerate}[i)]
		\item If $G' \sim G$, then $\cC(D,G') \sim \cC(D,G)$. \sk
\item If $\mathcal{C}$ is a linear code over $\fq$ equivalent to $\cC(D,G)$, then there exists a divisor $G'\sim G$ such that $G'$ and $D$ have disjoint supports and $\mathcal{C}=\cC(D,G')$.
		\item Moreover, for $z \in \cF^*$ such that $z(P_i) \neq 0, \infty$ for $i=1, \dots, n$ we have that $$\cC(D,G-(z))= {\bf x} \cdot \cC(D,G),$$ where ${\bf x}=(z(P_1), \dots , z(P_n)) \in (\fq^*)^n$. \sk
	\end{enumerate}
\end{proposition}

Part iii) is the geometric incarnation of a scaling: multiplying the evaluation
vectors by ${\bf x}=(z(P_1),\dots,z(P_n))$ amounts to replacing $G$ by
$G-(z)$, since $f\mapsto zf$ maps $\cL(G)$ onto $\cL(G-(z))$ when $(z)$ avoids
the $P_i$. Parts i) and ii) then show that the equivalence class of $\cC(D,G)$ is described by the linear equivalence class of $G$.

The hull, however, is \emph{not} an invariant of the equivalence class, and
this is the whole point of the problems \eqref{problem1} and \eqref{problem2}.
Before turning to the dual of an AG code, let us make this explicit. For
${\bf x}=(x_1, \dots, x_n) \in(\fq^\ast)^n$ and $c,y\in\fq^n$ we have
$\langle {\bf x}\cdot c,y\rangle=\langle c,{\bf x}\cdot y\rangle$, hence
$$
({\bf x}\cdot \cC)^{\perp}={\bf x}^{-1}\cdot \cC^{\perp},\qquad\text{where }{\bf x}^{-1}=(x_1^{-1},\dots,x_n^{-1}),
$$
and consequently
$$
\hull({\bf x}\cdot \cC)=({\bf x}\cdot \cC)\cap({\bf x}^{-1}\cdot \cC^{\perp}).
$$
The two codes appearing on the right-hand side are no longer related by the
\emph{same} scaling, so that their intersection may change with ${\bf x}$.

Multiplying the right-hand side by ${\bf x}^{-1}$ (a bijection of $\fq^n$,
hence compatible with intersections) gives the equivalent expression
$$
\dim_{\fq}\hull({\bf x}\cdot \cC)=\dim_{\fq}\big(({\bf x}^{2}\cdot \cC) \,\cap\, \cC^{\perp}\big),\qquad \text{where } {\bf x}^2=(x_1^2,\dots,x_n^2).
$$
In particular, the hull dimension of ${\bf x}\cdot \cC$ depends only on the
vector of squares ${\bf x}^2$. For $q$ odd, ${\bf x}^2$ ranges only over the vectors of non-zero squares of $\fq$, whereas for $q$ even every vector of $(\fq^\ast)^n$ is a vector of
squares, so that all scalings are effectively available.

\begin{example}
Let $q=5$ and $\cC=\langle(1,1)\rangle\subseteq\mathbb{F}_5^2$. Then $\cC^{\perp}=\langle(1,4)\rangle$ and $\cC\cap \cC^{\perp}=\{0\}$, so $\cC$ is LCD. For ${\bf x}=(1,2)$ one finds ${\bf x}\cdot \cC=\langle(1,2)\rangle$, and since $1^2+2^2=0$ in $\mathbb{F}_5$, this code is self-dual: $\dim_{\fq}\hull({\bf x}\cdot \cC)=1>0=\dim_{\fq}\hull(\cC)$. This example is consistent with the identity above: here ${\bf x}^2=(1,4)$, so ${\bf x}^2\cdot \cC=\langle(1,4)\rangle=\cC^{\perp}$, and the intersection ${\bf x}^2\cdot \cC\cap \cC^{\perp}$ has dimension~$1$.
\end{example}

\subsection{Duality for AG codes: the divisor of the dual code}\label{sec:dual}
The hull is defined through the dual code, so the next step is to describe
$\cC(D,G)^{\perp}$ in terms of divisors. The dual code $\cC^\perp$ of a linear code
$\cC$ over $\fq$ is the orthogonal complement of $\cC$ in $\fq^n$ with respect to the
standard Euclidean inner product of $\fq^n$. It was shown in \cite[Corollary~2.6]{S1988} that the dual of a linear AG code is also a linear AG code
formed with divisors that can be explicitly constructed. More precisely, we
have the following result.
\begin{proposition}
\label{AGdual}
The dual code of $\cC_{\cL}(D,G)$ is again a linear AG code; in fact,
\begin{equation}
	\cC(D,G)^\perp = \cC(D,D-G+(\eta)),
\end{equation}
where $\eta$ is a Weil differential in $\cF$ such that
$v_{P_i}(\eta)=-1$ and $\eta_{P_i}(1)=1$, for $i=1,\ldots, n$.
\end{proposition}

Throughout, we write $W:=(\eta)$ and $H:=D-G+W$, so that $\cC(D,G)^\perp=\cC(D,H)$. The \emph{code} $\cC(D,H)$ does not depend on the choice of $\eta$. Indeed, if $\eta'$ is another differential with the same properties, then $\eta'=h\eta$ for some $h\in\cF^\ast$ with $v_{P_i}(h)=0$ and $h(P_i)=1$ for every $i$, so that $W'=W+(h)$ and $H'=H+(h)$; since $h(P_i)=1$, multiplication by $h$ does not change evaluation vectors, and therefore $\cC(D,H')=\cC(D,H)$. Moreover, the identity $G+H=D+W$ shows that the passage $G\mapsto H$ is an involution up to the choice of $W$, which reflects the identity $(\cC^{\perp})^{\perp}=\cC$. The divisor $H$, however, does depend on $\eta$; this point is important for what follows and is discussed in Remark~\ref{rem:Wdependence}.

\subsection{The hull and the arithmetic of divisors}\label{sec:hullgcd}
We now reach the heart of the preliminaries: the link between the hull of an AG
code and the lattice structure of $\mathrm{Div}(\cF)$. Recall from the Introduction that the hull dimension lies in $\{0,\dots,\lfloor n/2\rfloor\}$, the value $0$ being attained exactly by the LCD codes and the value $n/2$ exactly by the self-dual codes.

Our computations of hull dimensions rest on the intersection of two
Riemann--Roch spaces, which is governed by the following lattice operations on
divisors. Let $\cF/\fq$ be an algebraic function field of genus $g$.
For two divisors $G_1$ and $G_2$ in $\mathrm{Div}(\cF)$, we define the
{\it greatest common divisor} of $G_1$ and $G_2$ by
	$$\gcd(G_1, G_2) = \sum_{P \in \mathcal P_\cF} \min\{ v_P(G_1), v_P(G_2) \} P,$$
	and  the {\it least common multiple} of $G_1$ and $G_2$ by
	$$\lcm(G_1,G_2) = \sum_{P \in \mathcal P_\cF} \max\{ v_P(G_1), v_P(G_2) \} P.$$
	The divisors $\gcd(G_1, G_2)$ and $\lcm(G_1, G_2)$ satisfy the following two fundamental properties:
\begin{equation}\label{basicP}
\cL(G_1) \cap \cL(G_2) = \cL(\gcd(G_1, G_2)) \quad \mbox{and} \quad \gcd(G_1, G_2)+\lcm(G_1, G_2) = G_1 + G_2.
\end{equation}
\begin{remark}
Two immediate by-products of \eqref{basicP} will be used without further
mention. First, taking degrees,
$\deg(\gcd(G_1,G_2))+\deg(\lcm(G_1,G_2))=\deg (G_1)+\deg (G_2)$. Second, since
$\gcd(G_1,G_2)\le G_i\le\lcm(G_1,G_2)$ for $i=1,2$, we have
$$
\cC(D,\gcd(G_1,G_2))\subseteq \cC(D,G_1)\cap \cC(D,G_2)
\qquad\text{and}\qquad
\cL(G_1)+\cL(G_2)\subseteq \cL(\lcm(G_1,G_2)).
$$
The first inclusion shows that $\cC(D,\gcd(G_1,G_2))$ is always contained in the intersection of the two codes; the whole question is when it is
an equality.
\end{remark}

The following theorem of \cite{CLM2024} is the starting point of our work. It
identifies the hull of an AG code with another AG code, the one attached to the
greatest common divisor of $G$ and the divisor $H$ describing the dual code, as
soon as this gcd is non-special.

\begin{theorem}\cite[Theorem 2, Corollary 3]{CLM2024}\label{hull nonspecial}
	Let \(G\), \(H\) and \(D = \sum_{i = 1}^{n} P_i\) be divisors of a function field \(\cF\) such that \(P_i \neq P_j\) for \(i \neq j\), and such that \(G\) and \(H\) each have support disjoint from that of \(D\). Assume that \(G + H - D\) is a canonical divisor such that
	\[
	\cC(D, G)^{\perp} = \cC(D, H).
	\]
	If \(\gcd(G, H)\) is non-special, then
	\[
	\operatorname{Hull}(\cC(D, G)) = \cC(D, \gcd(G, H)).
	\]
	We also have
	\begin{enumerate}[i)]
		\item If $\deg(\gcd(G, H)) <\deg(D)$, then \( \dim \hull \cC(D, G)=\ell(\gcd(G, H))\).
		\item If \(\deg( \gcd(G, H) )= g - 1\), then \(\cC(D, G)\) is LCD.
		\item If \(G \leq H\), then \(\cC(D, G)\) is self-orthogonal. If equality holds, then the code is self-dual.
	\end{enumerate}
\end{theorem}

Three situations are singled out. If $\deg(\gcd(G,H))<n$, the dimension formula $k=\ell(\cdot)-\ell(\cdot-D)$ of \eqref{dimcode} reduces to $\ell(\gcd(G,H))$, because $\ell(\gcd(G,H)-D)=0$; hence $\dim_{\fq}\hull \cC(D,G)=\ell(\gcd(G,H))=\deg(\gcd(G,H))+1-g$ by non-speciality. The LCD case corresponds to a non-special $\gcd(G,H)$ of degree $g-1$, whose Riemann--Roch space is $\{0\}$ by Lemma~\ref{lem:nonspecial}\,ii). Finally, $G\le H$ means $\gcd(G,H)=G$, that is, $\cC(D,G)\subseteq \cC(D,H)=\cC(D,G)^{\perp}$.

In the language of the present paper, Theorem~\ref{hull nonspecial} says that,
under non-speciality of the $\gcd$, the hull dimension is the \emph{degree} of the common
part of $G$ and $H$, up to the constant $1-g$:
$$\dim_{\fq}\hull \cC(D,G)=\deg(\gcd(G,H))+1-g.$$
This is the identity that converts questions about hulls into questions about
the lattice $\mathrm{Div}(\cF)$, and on which Section~\ref{sec:main} is built.

The theorem above requires non-speciality. To understand what happens in
general, and later to control the obstruction, we need a finer description of
the intersection of two AG codes. We now recall \cite[Theorem~3.1]{MS2026} in the form needed here.
It measures how far the intersection of two AG codes is from the AG code attached to the gcd. Since \cite{MS2026} is a preprint, we include a short proof for the reader's convenience.
\begin{theorem}\label{shaska}
Let $\cF/\fq$ be a function field of genus $g$. Let $D, G_1, G_2$ be divisors of $\cF$ such that $\supp(D) \cap \supp(G_1)=\supp(D) \cap \supp(G_2)=\emptyset$. Suppose that
$2g-2<\deg(G_i) < \deg(D)$ for $i=1, 2$ and consider the algebraic geometry codes $\cC(D, G_1)$ and $\cC(D, G_2)$. Then there is a canonical isomorphism of $\fq$-vector spaces
\begin{equation*}
\frac{\cC(D,G_1) \cap \cC(D,G_2) }{\cC( D, \gcd(G_1, G_2))}\,  \cong \, (\cL(G_1)+\cL(G_2)) \cap \cL( \lcm(G_1, G_2)-D).
\end{equation*}
Consequently,
\begin{align*}
\dim_{\fq} \cC(D,G_1) \cap \cC(D,G_2)&= \ell(\gcd(G_1, G_2))\\
&+ \dim_{\fq} ((\cL(G_1)+\cL(G_2)) \cap \cL( \lcm(G_1, G_2)-D) ).
\end{align*}
\end{theorem}
\begin{proof}
Since $\deg(G_i)<n$, the evaluation map is injective on $\cL(G_i)$, $i=1,2$. Hence every $c\in\cC(D,G_1)\cap\cC(D,G_2)$ is of the form $c=\mathrm{ev}_D(f_1)=\mathrm{ev}_D(f_2)$ for \emph{unique} $f_1\in\cL(G_1)$ and $f_2\in\cL(G_2)$, and we may define the $\fq$-linear map $\varphi(c):=f_1-f_2$. The function $f_1-f_2$ lies in $\cL(G_1)+\cL(G_2)\subseteq\cL(\lcm(G_1,G_2))$ and vanishes at every $P_i$, so it lies in $\cL(\lcm(G_1,G_2)-D)$. The kernel of $\varphi$ consists of the codewords with $f_1=f_2\in\cL(G_1)\cap\cL(G_2)=\cL(\gcd(G_1,G_2))$, that is, $\ker\varphi=\cC(D,\gcd(G_1,G_2))$. Finally, $\varphi$ is onto: if $h=f_1-f_2$ with $f_i\in\cL(G_i)$ lies in $\cL(\lcm(G_1,G_2)-D)$, then $h(P_i)=0$ for all $i$, so $c:=\mathrm{ev}_D(f_1)=\mathrm{ev}_D(f_2)$ belongs to $\cC(D,G_1)\cap\cC(D,G_2)$ and $\varphi(c)=h$.
\end{proof}

The ``consequently'' follows from the isomorphism since $\deg(\gcd(G_1,G_2))\le\deg (G_1)<\deg (D)$, so that $\ell(\gcd(G_1,G_2)-D)=0$ and $\dim_{\fq} \cC(D,\gcd(G_1,G_2))=\ell(\gcd(G_1,G_2))$. The space on the right-hand side is the \emph{defect} by which the intersection of the two codes exceeds $\cC(D,\gcd(G_1,G_2))$.

In the particular case $\cC(D, G_2)=\cC(D, G_1)^\perp$ of the last theorem, we obtain the following corollary.

\begin{corollary}\label{shaskadual} Fix the notation as in Theorem \ref{shaska}. Let $G$ be a divisor such that $2g-2 < \deg(G) < n$, and let $\eta$ be a Weil differential in $\cF$ such that
$v_{P_i}(\eta)=-1$ and $\eta_{P_i}(1)=1$, for $i=1,\ldots,n$. Let $W=(\eta)$ and $H=D-G+W$, so that $\cC(D, G)^\perp=\cC(D, H)$.
Then
\begin{align*}
	\dim_{\fq} \hull \cC(D,G) & =\ell(\gcd(G, H))\\
	&+ \dim_{\fq} ((\cL(G)+\cL(H)) \cap \cL(W-\gcd(G, H))).
\end{align*}
\end{corollary}
\begin{proof} 
Use the fact that $\gcd(G, H)+\lcm(G, H)=G+H=D+W$, and apply Theorem~\ref{shaska} with $G_1=G$ and $G_2=H$.
\end{proof}

Corollary~\ref{shaskadual} makes visible the role of the non-speciality hypothesis in Theorem~\ref{hull nonspecial}. If $\gcd(G,H)$ is non-special, then $\ell(W-\gcd(G,H))=0$, the second summand vanishes and we recover $\dim_{\fq}\hull \cC(D,G)=\ell(\gcd(G,H))$. Without this hypothesis, the second summand may be positive, and it is the obstruction that has to be controlled in Section~\ref{sec:general}. Note also that the hypothesis $2g-2<\deg (G)<n$ is compatible with Theorem~\ref{shaska}, since $\deg (H)=n+2g-2-\deg (G)$ lies in the same range.

The second summand in Corollary~\ref{shaskadual} is a subspace of
$\cL(W-\gcd(G,H))$, whose dimension is, by definition, the index of speciality
$i(\gcd(G,H))$. Hence the defect is bounded by $i(\gcd(G,H))$ and vanishes as
soon as this index does. This gives a quantitative version of the dichotomy
``non-special $\Rightarrow$ the hull is $\cC(D,\gcd(G,H))$'' which underlies
Sections~\ref{sec:main} and~\ref{sec:general}.

\subsection{The maximal hull dimension of a code: Chen's problem in the AG setting}\label{sec:maxhull}
We close the preliminaries by recalling the problem that motivates the paper,
now stated for AG codes. In \cite{C2023}, Chen proposed to investigate the {\it maximum Euclidean hull dimension} of a code $\cC$, that is, to investigate the behaviour of the hull dimension under equivalence. More precisely, Chen considered
\begin{equation}
\maxEhull(\cC) := \max_{\mathbf{x} \in (\mathbb{F}_q^*)^n} \dim_{\fq}  \left( (\mathbf{x} \cdot \cC) \cap (\mathbf{x} \cdot \cC)^\perp \right).
\end{equation}
In the case of a function field $\cF/\fq$ with pairwise distinct rational places $P_1, \dots, P_n$, the Approximation Theorem \cite[Theorem~1.3.1]{S2009} shows that any vector $\mathbf{x} \in (\mathbb{F}_q^*)^n$ can be written as
$$\mathbf{x} =(z(P_1), \dots, z(P_n))$$
for some function $z\in\cF^*$. In particular, for AG codes, where $D=\sum_{i=1}^n P_i$ and $W=(\eta)$ is a canonical divisor such that
$v_{P_i}(\eta)=-1$ and $\eta_{P_i}(1)=1$, for $i=1,\ldots,n$, the investigation of the maximum Euclidean hull dimension of an AG code $\cC(D,G)$ translates into
\begin{align*}
\maxEhull(\cC(D, G))& = \max_{\mathbf{x} \in (\mathbb{F}_q^*)^n} \dim_{\fq}  \left( (\mathbf{x} \cdot \cC(D, G)) \cap (\mathbf{x} \cdot \cC(D, G))^\perp \right)\\
&=  \max_{\substack{z \in \cF^*,\\ \supp((z))\cap \supp(D)=\emptyset}} \dim_{\fq}  \left( ( \cC(D, G-(z))) \cap (\cC(D, D-G+W+(z))) \right)
\end{align*}
by Proposition \ref{prop2210}.

In other words, $\maxEhull(\cC(D,G))$ is the largest value of $\dim\hull$ on the equivalence class of $\cC(D,G)$; the pair of divisors $(G-(z),\,D-G+W+(z))=(G-(z),\,H+(z))$ is obtained from $(G,H)$ by the twist $L=-(z)$, which is exactly the mechanism studied in Section~\ref{sec:main}.

\section{Hull variation under a non-speciality condition on the greatest common divisor}\label{sec:main}

The dictionary of Section~\ref{sec:preliminaries} reduces the hull of $\cC(D,G)$ to the arithmetic of the pair of divisors $(G,H)$, where $H=D-G+W$ describes the dual code. A twist $G\mapsto G+L$ acts on this pair by $(G,H)\mapsto(G+L,H-L)$, so that the sum $G+H=D+W$ is \emph{invariant}. The hull being governed by the common part $\gcd(G,H)$, the variation of its dimension is controlled by how the twist modifies $\gcd(G,H)$. This is the guiding principle of the present section.

The plan is as follows. After fixing the standing assumptions, we compute the variation exactly and give a sharp upper bound for the hull dimension (Section~\ref{sec:exact}); we then specialise to principal twists, which correspond to scaling equivalence, and apply the results to Chen's maximal hull dimension (Section~\ref{sec:principal}); finally, we reverse the point of view and \emph{construct} perturbations of $G$ achieving prescribed hull dimensions (Section~\ref{sec:raiselower}).

We begin by establishing the following notation and hypotheses, which will be assumed throughout this section.

\begin{assumption}\label{ass:main}
	Let $\mathcal{F}/\mathbb{F}_q$ be a function field of genus $g>0$ over the finite field $\mathbb{F}_q$.
	\begin{enumerate}[i)]
		\item Fix $P_1, \dots, P_n$ pairwise distinct rational places in $\mathcal{F}$, set $D=P_1+\cdots+P_n$, and let $G$ be a divisor such that $\operatorname{supp}(G)\cap \operatorname{supp}(D)=\emptyset$.
		\item Assume that $2g-2<\deg(D)=n$.
		\item $W=(\eta)$ is a fixed canonical divisor satisfying $v_{P_i}(\eta)=-1$ and $\eta_{P_i}(1)=1$ for all $i=1,\ldots, n$; define $H:=D-G+W$.
		\item Assume that $\gcd(G,H)$ is a non-special divisor.
	\end{enumerate}
\end{assumption}

\begin{remark}[Dependence on the choice of $W$]\label{rem:Wdependence}
The code $\cC(D,G)^\perp=\cC(D,H)$ does not depend on the choice of the differential $\eta$ in Assumption~\ref{ass:main}\,iii), but the divisors $W$ and $H$ do: another admissible choice is $\eta'=h\eta$ with $h\in\cF^\ast$ and $h(P_i)=1$ for all $i$, which replaces $(W,H)$ by $(W+(h),H+(h))$. Consequently, the divisors $G-H$ and $\gcd(G,H)$, the non-speciality condition iv), the divisors $A_W$ and $R_W$ of Proposition~\ref{thm:upper-bound-hull}, and the sets $\mathcal{T}_D(G)$, $U_D(G)$ introduced below all depend on the chosen $W$; for instance, $R_{W+(h)}$ differs from $R_W$ at every zero or pole of $h$ of odd order. Throughout Sections~\ref{sec:main}--\ref{sec:general}, $W$ is fixed once and for all, and all statements are to be understood for this fixed $W$. On the other hand, $\dim_{\fq}\hull\cC(D,G)$ is an invariant of the code. Hence, if two admissible choices $W$ and $W'=W+(h)$ both satisfy iv), then \eqref{eq_closedformula} below yields
$$
\deg(|G-H|)=\deg(|G-H-(h)|).
$$
In particular, the bound \eqref{eq_bound_RW} below holds for \emph{every} admissible $W$ satisfying iv), so that the best bound is obtained by choosing such a $W$ with $\deg(R_W)$ as large as possible. Determining this optimal choice in general is left open.
\end{remark}

Under Assumption~\ref{ass:main}, Proposition~\ref{AGdual} and Theorem~\ref{hull nonspecial} give
$$
\cC(D,G)^{\perp}=\cC(D,H) \quad \text{and} \quad
\hull\big(\cC(D,G)\big)=\cC\big(D,\gcd(G,H)\big).
$$
We now fix some notation and present some elementary identities that are used repeatedly in this section. Given a divisor $A=\sum_{P\in \cP_{\cF}}v_P(A)P$, let  $$|A|:=\sum_{P\in \cP_{\cF}}|v_P(A)|P\quad \text{and}\quad A_+=\sum_ {\substack{P \in \cP_\cF \\ v_P(A) >0}} v_P(A)P.$$
Note that we can always write  $A=A_+ - (-A)_+$.

For integers $a, b$ we have $\min\{a,b\}=\frac12\big((a+b)-|a-b|\big)$. Applying this identity place by place to divisors $A, B$ yields
$
\gcd(A,B)=\tfrac12\big(A+B-|A-B|\big).
$
Taking degrees on both sides of the equality, we obtain
\begin{equation}\label{gcddegree}
	\deg(\gcd(A,B))=\tfrac12\big(\deg (A)+\deg (B)-\deg(|A-B|)\big).
\end{equation}
Moreover, $\big||A|-|B|\big|\le|A-B|$ implies $\big|\deg(|A|)-\deg(|B|)\big|\le\deg(|A-B|)$, and $|A+B|=|A|+|B|$ whenever $\operatorname{supp}(A)\cap\operatorname{supp}(B)=\emptyset$.

\subsection{An exact formula for the variation of the hull under a divisor twist}\label{sec:exact}

The following result determines, under the non-speciality condition, the hull variation of the code $\cC(D,G+L)$, which is obtained by perturbing the divisor $G$ by a divisor $L$, with respect to the code $\cC(D,G)$. 

\begin{theorem}\label{prop_perturbation_hull}
	Let $D,G$ and $H$ be the divisors defined in Assumption \ref{ass:main}.
	Let $L$ be a divisor of $\mathcal{F}$ such that
	\begin{enumerate}[i)]
		\item $\supp(L)\cap\supp(D)=\emptyset$, and
		\item $\gcd(G+L,H-L)$ is non-special.
	\end{enumerate}
	Then
	\begin{equation}\label{difference}
		\dim_{\fq} \hull (\cC(D,G+L))-\dim_{\fq} \hull (\cC(D,G))=\frac{1}{2}(\deg( |G-H|)-\deg (|G-H+2L|)).
	\end{equation}
	In particular,
	\begin{enumerate}[a)]
		\item $|\dim_{\fq} \hull(\cC(D,G+L))-\dim_{\fq} \hull(\cC(D,G))|\leq \deg( |L|), $ and
		\item if $\supp(L)\cap \supp(G-H)=\emptyset$, then $$\dim_{\fq} \hull(\cC(D,G+L))=\dim_{\fq} \hull(\cC(D,G))-\deg (|L|).$$
	\end{enumerate}
\end{theorem}
\begin{proof} 

We compute the degrees of $\gcd(G,H)$ and of $\gcd(G+L,H-L)$ by applying \eqref{gcddegree}; we identify $\hull(\cC(D,G+L))$ with an AG code via Theorem~\ref{hull nonspecial}; and we count its dimension by the Riemann--Roch theorem, which is applicable thanks to non-speciality.

Since $\deg(G+H)=\deg(D+W)=n+2g-2$, \eqref{gcddegree} gives
	\begin{equation}\label{deggcdGH}
\begin{split}
\bullet\ \deg(\gcd(G,H)) &= \frac{1}{2}\bigl(\deg(G+H)-\deg(|G-H|)\bigr)\\
&= \frac{1}{2}\bigl(n+2g-2-\deg(|G-H|)\bigr).
\end{split}
\end{equation} and
	\begin{equation}\label{deggcdGHL}
\begin{split}
\bullet\ \deg(\gcd(G+L,H-L)) &= \frac{1}{2}\bigl(\deg(G+H)-\deg(|G-H+2L|)\bigr)\\
&= \frac{1}{2}\bigl(n+2g-2-\deg(|G-H+2L|)\bigr).
\end{split}
\end{equation}
	Moreover, since $2g-2<n$, we conclude that
	$$
	\deg(\gcd(G,H))< n\quad \text{and}\quad \deg(\gcd(G+L,H-L))< n.
	$$
 On the other hand, since $\gcd(G+L,H-L)$ is non-special, from Proposition~\ref{AGdual} and Theorem~\ref{hull nonspecial} we have
	\begin{align*}
		\hull(\cC(D,G+L))&=\cC(D,G+L)\cap \cC(D,G+L)^\perp\\
		&=\cC(D,G+L)\cap \cC(D, D-(G+L)+W) \\
		&=\cC(D,G+L)\cap \cC(D,H-L)\\
		&=\cC(D, \gcd(G+L,H-L)).
	\end{align*}

 Since $\deg(\gcd(G+L,H-L))<n$, Theorem~\ref{hull nonspecial}\,i) and the Riemann--Roch theorem (which applies by non-speciality) give
	
	\begin{equation}\label{eq:hull_twist}
\begin{aligned}
  \dim_{\fq} \hull (\cC(D,G+L)) &= \ell(\gcd(G+L,H-L))\\
  &= \deg(\gcd(G+L,H-L))+1-g\\
  &= \frac{1}{2}\bigl(n-\deg(|G-H+2L|)\bigr).
\end{aligned}
\end{equation}
	Analogously, 
	\begin{equation} \label{hullnotwist}
	\dim_{\fq} \hull(\cC(D,G))=\frac{1}{2}(n-\deg(|G-H|)).
	\end{equation}
	Thus we get
	\begin{equation*}
			\dim_{\fq} \hull (\cC(D,G+L))-\dim_{\fq} \hull (\cC(D,G))=\frac{1}{2}(\deg(|G-H|)-\deg(|G-H+2L|)).
	\end{equation*}
	This concludes the first part of the proof.

	 For the second part of the proof, note that $$|\deg(|A|)-\deg(|B|)|\leq \deg (|A-B|)$$ for any divisors $A$ and $B$ in $\cF$. Therefore
	\begin{align*}
	|\dim_{\fq} \hull (\cC(D,G+L))-\dim_{\fq} \hull (\cC(D,G))|&=\frac{1}{2}|\deg(|G-H|)-\deg(|G-H+2L|)|\\
	&\leq \deg(|L|).
	\end{align*}
	In particular, if $\supp(L)\cap \supp(G-H)=\emptyset$, then $$\deg(|G-H+2L|)=\deg(|G-H|)+2\deg(|L|)$$ and
	$$
	\dim_{\fq} \hull (\cC(D,G+L))=\dim_{\fq} \hull (\cC(D,G))- \deg(|L|).
	$$
\end{proof}

\begin{remark}\label{rem:towardaway}
Now we put Theorem~\ref{prop_perturbation_hull} into perspective.
Since $G-H=2G-D-W$, from the proof and under Assumption~\ref{ass:main},
\begin{equation}\label{eq_closedformula}
\dim_{\fq}\hull \cC(D,G)=\tfrac12\big(n-\deg(|2G-D-W|)\big)=\tfrac12\big(n-\deg(|G-H|)\big).
\end{equation}
The hull dimension is thus governed by the ``distance'' $\deg(|2G-D-W|)$ between $2G$ and the divisor $D+W$. It reaches its maximal value $n/2$ if and only if $G=H$, which is the self-dual case of Theorem~\ref{hull nonspecial}\,iii).
Note that, by b), the bound in a) is attained, so it cannot be improved without further hypotheses. The bound in a) is also attained by \emph{increases} of the hull: by Theorem~\ref{hull+} below, for suitable twists $L$ the hull grows by exactly $\deg(|L|)$.
\end{remark}

In the next result we express the hull dimension of $\cC(D,G)$ in terms of two divisors determined by $D$ and $W$, and we obtain an upper bound for it.

\begin{proposition}\label{thm:upper-bound-hull}
	Let $D$, $G$, $H$ and $W$ be as in Assumption~\ref{ass:main}. Write $D+W=2A_W+R_W$, where 
	$$
	A_W=\sum_{P\in\cP_\cF\setminus\supp(D)}\Big\lfloor\frac{v_P(W)}{2}\Big\rfloor P
	\qquad\text{and}\qquad
	R_W=\sum_{\substack{P\in\cP_\cF\setminus\supp(D)\\ v_P(W)\ \text{odd}}}P.
	$$
	Then
	\begin{equation}\label{boxformula}
		\dim_{\fq}\hull\big(\cC(D,G)\big)=\frac12\big(n-\deg( R_W)\big)-\deg(\big(G-A_W-R_W\big)_{+})-\deg(\big(A_W-G\big)_{+}).
	\end{equation}
	In particular,
	\begin{equation}\label{eq_bound_RW}
	\dim_{\fq}\hull\big(\cC(D,G)\big)\le\frac12\big(n-\deg (R_W)\big),
	\end{equation}
	with equality if and only if $A_W\le G\le A_W+R_W$, or equivalently, if and only if $|v_P(G-H)|\le 1$ for all $P\in\cP_\cF$.
\end{proposition}

\begin{proof}
	By \eqref{eq_closedformula}, we have
	$$
	\dim_{\fq}\hull\big(\cC(D,G)\big)=\frac12\big(n-\deg(|G-H|)\big).
	$$
	Since $H+G=D+W=2A_W+R_W$, we have $G-H=2(G-A_W)-R_W$. Moreover, for all integers $a\in \Z$ and $b\in \{0,1\}$ one checks directly that
	\begin{equation}\label{localidentity}
	|2a-b|=b+2\max\{a-b,0\}+2\max\{-a,0\}.
	\end{equation}
	For  $P\in\cP_\cF$, let $a_P=v_P(G-A_W)\in\Z$ and $b_P=v_P(R_W)\in\{0,1\}$. This yields
	\begin{align*}
	\deg (|G-H|)&=\deg( |2(G-A_W)-R_W|)\\
	&= \sum_{P \in {\cP}_{\cF}} |2a_P-b_P| \deg (P)\\
	&=\sum_{P \in {\cP}_{\cF}} ( b_P+2\max\{a_P-b_P,0\}+2\max\{-a_P,0\})\deg (P)\\
	&=\deg ( R_W)+2\deg(\big(G-A_W-R_W\big)_{+})+2\deg(\big(A_W-G\big)_{+}),
	\end{align*}
which yields \eqref{boxformula} and the upper bound. We have equality in the upper bound if and only if $(G-A_W-R_W)_{+}=0$ and $(A_W-G)_{+}=0$, that is, $A_W\le G\le A_W+R_W$. 

Finally, by \eqref{localidentity}, $\max\{a_P-b_P,0\}=\max\{-a_P,0\}=0$ if and only if $|2a_P-b_P|=b_P$. Since $|2a_P-b_P|\equiv b_P\bmod{2}$ and $b_P\in\{0,1\}$, this is equivalent to $|2a_P-b_P|\le 1$, that is, $|v_P(G-H)|\le 1$. Hence $A_W\le G\le A_W+R_W$ if and only if $|v_P(G-H)|\le 1$ for all $P\in\cP_\cF$.
\end{proof}

\begin{remark}\label{rem:upperbound}
	(1) Let
	$$
	\mathcal{T}_D(G):=\{L\in\mathrm{Div}(\cF) : \supp(L)\cap\supp(D)=\emptyset \text{ and } \gcd(G+L,H-L)\ \text{is non-special}\}
	$$
	be the set of twists allowed in Theorem~\ref{prop_perturbation_hull}. If $L\in\mathcal{T}_D(G)$, then, since $H-L=D-(G+L)+W$, the divisors $D$, $G+L$, $H-L$ and $W$ satisfy Assumption~\ref{ass:main} with the same canonical divisor $W$. Hence Proposition~\ref{thm:upper-bound-hull} applies to $\cC(D,G+L)$, and therefore
	\begin{equation}\label{eq_bound_max_L}
	\max_{L\in\mathcal{T}_D(G)}\dim_{\fq}\hull\big(\cC(D,G+L)\big)\le\frac12\big(n-\deg( R_W)\big).
	\end{equation}
	We will show in Theorem~\ref{hull+} that, for any divisors $D$, $G$, $H$ and $W$ satisfying Assumption~\ref{ass:main}, there always exists $L\in\mathcal{T}_D(G)$ such that $\dim_{\fq}\hull(\cC(D,G+L))=\frac12(n-\deg (R_W))$. Thus the bound of Proposition~\ref{thm:upper-bound-hull} is sharp, and the inequality above is in fact an equality.
	
	\smallskip
	(2) Every linear $[n, k]$ code $\cC$ satisfies the trivial bound $\dim_{\fq}\hull(\cC)\le\min\{k,n-k\}$. Combining this with Proposition~\ref{thm:upper-bound-hull}, we obtain
	$$
	\dim_{\fq}\hull\big(\cC(D,G)\big)\le\min\Big\{k,\ n-k,\ \tfrac12\big(n-\deg (R_W)\big)\Big\}.
	$$
	Therefore, Proposition~\ref{thm:upper-bound-hull} improves the trivial bound if and only if $|2k-n|<\deg (R_W)$, that is,
	$$
	\tfrac12\big(n-\deg (R_W)\big)<k<\tfrac12\big(n+\deg( R_W)\big).
	$$
\end{remark}

\subsection{Principal twists: scaling equivalence and Chen's problem}\label{sec:principal}

Specialising $L$ to a principal divisor brings us back to the hull-variation problem \eqref{problem1} stated in the Introduction, and to Chen's maximal hull dimension of Section~\ref{sec:maxhull}. Set
$$
U_D=\{z\in\cF^{*} : \supp((z))\cap \supp (D)=\emptyset\}
$$
and 
$$
{\bf x}_z=(z(P_1),\dots,z(P_n))\in(\fq^{*})^n \ \text{ for } z\in U_D .
$$
We also write
$$
U_D(G)=\{z\in U_D : \gcd(G-(z),H+(z))\ \text{is non-special}\}.
$$
By Proposition~\ref{prop2210}\, iii), ${\bf x}_z\cdot\cC(D,G)=\cC(D,G-(z))$, and its dual is $\cC(D,H+(z))$ with $H+(z)=D-(G-(z))+W$; that is, the dual of the twisted code is described by the same canonical divisor $W$. By the Approximation Theorem, every ${\bf x}\in(\fq^*)^n$ is of the form ${\bf x}_z$ for some $z\in U_D$, so that
$$
\maxEhull(\cC(D,G))=\max_{z\in U_D}\dim_{\fq}\hull({\bf x}_z\cdot\cC(D,G)).
$$
As a consequence of Theorem~\ref{prop_perturbation_hull}, we first obtain the hull variation of the code ${\bf x}_z\cdot\cC(D,G)$ with respect to $\cC(D,G)$.

\begin{corollary}\label{cor:principal}
	Let $D, G, H$ be the divisors defined in Assumption \ref{ass:main}, and let $z\in U_D(G)\setminus\fq$. Then
	\begin{equation*}
		\begin{split}
			\dim_{\fq} \hull({\bf x}_z\cdot \cC(D,G))-\dim_{\fq} \hull(\cC(D,G))=\frac{1}{2}(\deg(|G-H|)-\deg(|G-H-2(z)|)).
		\end{split}
	\end{equation*}
	In particular,
	$$
	|\dim_{\fq} \hull({\bf x}_z\cdot \cC(D,G))-\dim_{\fq} \hull(\cC(D,G))|\leq 2[\cF:\fq(z)].
	$$
\end{corollary}
\begin{proof}
	The result follows immediately from Theorem \ref{prop_perturbation_hull} by setting $L=-(z)$. Note that $|L|=|(z)|=(z)_0+(z)_\infty$ and $\deg((z)_0)=\deg((z)_\infty)=[\mathcal{F}:\mathbb{F}_q(z)]$.
\end{proof}

In the next result, we give an upper bound for the maximal hull dimension of the codes ${\bf x}_z\cdot\cC(D,G)$ with $z\in U_D(G)$.

\begin{proposition}\label{prop:chenupper}
	Let $D$, $G$, $H$ and $W$ be as in Assumption~\ref{ass:main}, and let
	$$
	R_W=\sum_{\substack{P\in\cP_\cF\setminus\supp(D)\\ v_P(W)\ \text{odd}}}P.
	$$
	 Then
	$$
	\max_{z\in U_D(G)}\dim_{\fq}\hull({\bf x}_z\cdot\cC(D,G))\le \tfrac12\big(n-\deg (R_W)\big).
	$$
\end{proposition}
\begin{proof}
	Let $z\in U_D(G)$. Then ${\bf x}_z\cdot\cC(D,G)=\cC(D,G-(z))$, and since $z\in U_D(G)$, the divisor $L=-(z)$ belongs to $\mathcal{T}_D(G)$. Hence, by Remark~\ref{rem:upperbound}\,(1), the divisors $D$, $G-(z)$, $H+(z)$ and $W$ satisfy Assumption~\ref{ass:main}, and Proposition~\ref{thm:upper-bound-hull} gives $\dim_{\fq}\hull({\bf x}_z\cdot\cC(D,G))\le\frac12(n-\deg (R_W))$. The result follows by taking the maximum over $z\in U_D(G)$.
\end{proof}

Note that $\maxEhull(\cC(D,G))$ is the maximum over $U_D$, whereas Proposition~\ref{prop:chenupper} only controls the maximum over $U_D(G)$. Moreover, by Remark~\ref{rem:upperbound}\,(1), the bound $\frac12(n-\deg( R_W))$ is always attained by some divisor $L\in\mathcal{T}_D(G)$, but this is no longer true when we restrict to principal divisors $L=-(z)$ with $z\in U_D(G)$. In the next result we characterise when equality holds in this case.

\begin{theorem}\label{thm:chenattain}
	Let $D$, $G$, $H$ and $W$ be as in Assumption~\ref{ass:main}, and let $A_W$ and $R_W$ be as in Proposition~\ref{thm:upper-bound-hull}, that is,
	 $$
	A_W=\sum_{P\in\cP_\cF\setminus\supp(D)}\Big\lfloor\frac{v_P(W)}{2}\Big\rfloor P
	\qquad\text{and}\qquad
	R_W=\sum_{\substack{P\in\cP_\cF\setminus\supp(D)\\ v_P(W)\ \text{odd}}}P.
	$$
	Then
	$$
	\max_{z\in U_D(G)}\dim_{\fq}\hull({\bf x}_z\cdot\cC(D,G))=\frac12\big(n-\deg( R_W)\big)
	$$
	if and only if $G\sim A_W+E$ for some divisor $E$ with $0\le E\le R_W$. In this case, the maximum is attained by any $z\in\cF^*$ with $(z)=G-A_W-E$, and ${\bf x}_z\cdot\cC(D,G)=\cC(D,A_W+E)$.
\end{theorem}
\begin{proof}
For $z\in U_D$ we write $G'=G-(z)$ and $H'=H+(z)=D-G'+W$, so that $${\bf x}_z\cdot\cC(D,G)=\cC(D,G')\, \text{ and } \, ({\bf x}_z\cdot\cC(D,G))^\perp=\cC(D,H').$$
	
Suppose that $G\sim A_W+E$ for some divisor $E$ with $0\le E\le R_W$, and let $z\in\cF^*$ with $(z)=G-A_W-E$. Since $G$, $A_W$ and $E$ have supports disjoint from $\supp(D)$, we have $z\in U_D$. Note also that $G'=A_W+E$ and $H'=D+W-G'=A_W+R_W-E$. Since $R_W$ has coefficients in $\{0,1\}$ and $0\le E\le R_W$, at each place $P$ at least one of $v_P(E)$ and $v_P(R_W-E)$ vanishes, so  
$$\gcd(G',H')=\gcd(A_W+E,A_W+R_W-E)=A_W.$$ 
We now prove that $A_W$ is non-special, which shows that $z\in U_D(G)$. Indeed, since $G+H=D+W=2A_W+R_W$, for every $P\in\cP_\cF$ we have 
	$$2\min\{v_P(G),v_P(H)\}\le v_P(G+H)=2v_P(A_W)+v_P(R_W).$$
	Since $v_P(R_W)\in\{0,1\}$, this gives $\min\{v_P(G),v_P(H)\}\le v_P(A_W)$, and hence $\gcd(G,H)\le A_W$. By hypothesis, $\gcd(G,H)$ is non-special, so $A_W$ is also non-special by Lemma~\ref{lem:nonspecial}\,iii). 
	
We have $z\in U_D(G)$, and the divisors $D$, $G'$, $H'$ and $W$ satisfy Assumption~\ref{ass:main}. Since $A_W\le G'\le A_W+R_W$, Proposition~\ref{thm:upper-bound-hull} gives $\dim_{\fq}\hull(\cC(D,G'))=\frac12(n-\deg (R_W))$, and by Proposition~\ref{prop:chenupper} this is the maximum.
	
Conversely, if some $z\in U_D(G)$ satisfies $\dim_{\fq}\hull(\cC(D,G'))=\frac12(n-\deg (R_W))$, then, since the divisors $D$, $G'$, $H'$ and $W$ satisfy Assumption~\ref{ass:main}, the equality case of Proposition~\ref{thm:upper-bound-hull} applied to $G'$ gives $A_W\le G'\le A_W+R_W$. Hence $E:=G'-A_W$ satisfies $0\le E\le R_W$ and $G=G'+(z)\sim A_W+E$.
\end{proof}

Note that Theorem~\ref{thm:chenattain} concerns the maximum over $U_D(G)$, not over $U_D$; it therefore does not determine $\maxEhull(\cC(D,G))$ in general, but it identifies exactly when the bound $\frac12(n-\deg(R_W))$ is reached by a scaling for which the gcd remains non-special.

The condition of Theorem~\ref{thm:chenattain} has a simple necessary consequence. Since $\deg(A_W)=\frac12\big(n+2g-2-\deg(R_W)\big)$, the relation $G\sim A_W+E$ with $0\le E\le R_W$ forces
$$
\tfrac12\big(n+2g-2-\deg(R_W)\big)\le\deg(G)\le\tfrac12\big(n+2g-2+\deg(R_W)\big).
$$
If moreover $2g-2<\deg(G)<n$, so that $k=\deg(G)+1-g$, this reads $|2k-n|\le\deg(R_W)$, in accordance with Remark~\ref{rem:upperbound}\,(2): only codes of dimension close to $n/2$ can reach the bound by a scaling.

In practice, the condition can be tested by checking, for each of the at most $2^{|\supp(R_W)|}$ divisors $E$ with $0\le E\le R_W$ and $\deg(E)=\deg(G)-\deg(A_W)$, whether $G-A_W-E$ is principal.

\subsection{Raising and lowering the hull dimension: prescribed perturbations of \texorpdfstring{$G$}{G}}\label{sec:raiselower}

Theorem~\ref{prop_perturbation_hull} is an \emph{a posteriori} formula: it computes the effect of a given twist $L$. In applications, one rather wishes to \emph{choose} $L$ so as to obtain a prescribed effect. In the next two theorems, as a further consequence of Theorem~\ref{prop_perturbation_hull}, we provide, under mild hypotheses, a way to construct AG codes whose hull dimension is larger or smaller than $\dim_{\fq} \hull(\cC(D,G))$. The first one raises the hull: starting from a code whose hull has dimension $\ell$, it produces codes whose hull has dimension $\ell+\delta$, for $\delta$ ranging over an explicit set of integers.

  \begin{theorem}[Raising the hull dimension] \label{hull+}
 	Let $D$, $G$, $H$ be divisors as in Assumption \ref{ass:main}. Define
 	$$
 	T_{+}=\supp\big((H-G)_{+}\big)
 	\quad\text{and}
 	\quad
 	T_{-}=\supp\big((G-H)_{+}\big).
 	$$
 	For $P\in T_{+}$, let 
 	$
 	0\le t_P\le\left\lfloor\frac{v_P(H)-v_P(G)}{2}\right\rfloor,
 	$
 	and for  $P\in T_{-}$, let
 	$
 	0\le s_P\le\left\lfloor\frac{v_P(G)-v_P(H)}{2}\right\rfloor.
 	$
 	Set
 	$$
 	L=\sum_{P\in T_{+}}t_P\,P-\sum_{P\in T_{-}}s_P\,P .
 	$$
 	Then $\supp(L)\cap\supp(D)=\emptyset$, the divisor $\gcd(G+L,H-L)$ is non-special, and
 	 \begin{equation}
 	\dim_{\mathbb{F}_q}\operatorname{Hull}\big(\cC(D,G+L)\big)=\dim_{\mathbb{F}_q}\operatorname{Hull}\big(\cC(D,G)\big)+\sum_{P\in T_{+}}t_P\deg(P)+\sum_{P\in T_{-}}s_P\deg(P).
 	\end{equation}
 	In particular, taking $t_P=\floor*{\frac{v_P(H)-v_P(G)}{2}}$ and $s_P=\floor*{\frac{v_P(G)-v_P(H)}{2}}$ for $P\in T_+$ and $P\in T_-$, respectively, we have	
	\[
 	\dim_{\mathbb{F}_q}\operatorname{Hull}\big(\cC(D,G+L)\big)=\frac{1}{2}\left(n-\deg (R_W )\right),
 	\]
	where $R_W=\sum_{\substack{P\in\cP_\cF\setminus\operatorname{supp}(D)\\ v_P(W)\ \text{odd}}} P$.
 \end{theorem}
 
 \begin{proof}
 First, note that for $P_i\in\operatorname{supp}(D)$ we have $v_{P_i}(G)=v_{P_i}(H)=0$. Hence $P_i\notin T_{+}\cup T_{-}$, and so $\operatorname{supp}(L)\cap\operatorname{supp}(D)=\emptyset$.

 On the other hand, let $P\in\cP_\cF$. We consider three cases.
 	\begin{itemize}
 		\item If $P\notin T_{+}\cup T_{-}$, then $v_P(L)=0$, so
 		\[
 		\min\{v_P(G+L),v_P(H-L)\}=\min\{v_P(G),v_P(H)\}.
 		\]
 		\item If $P\in T_{+}$, then $2t_P\le v_P(H)-v_P(G)$, that is, $v_P(G)+t_P\le v_P(H)-t_P$. Hence
 		\[
 		\min\{v_P(G)+t_P,\ v_P(H)-t_P\}=v_P(G)+t_P=\min\{v_P(G),v_P(H)\}+t_P .
 		\]
 		\item If $P\in T_{-}$, then $2s_P\le v_P(G)-v_P(H)$, that is, $v_P(H)+s_P\le v_P(G)-s_P$. Hence
 		\[
 		\min\{v_P(G)-s_P,\ v_P(H)+s_P\}=v_P(H)+s_P=\min\{v_P(G),v_P(H)\}+s_P .
 		\]
 	\end{itemize}
 	Therefore
 	\[
 	\gcd(G+L,H-L)=\gcd(G,H)+\sum_{P\in T_{+}}t_P\,P+\sum_{P\in T_{-}}s_P\,P=\gcd(G,H)+|L| .
 	\]
 	Since $|L|\ge 0$, we have $\gcd(G,H)\le\gcd(G+L,H-L)$. Because $\gcd(G,H)$ is non-special, $\gcd(G+L,H-L)$ is non-special as well by Lemma~\ref{lem:nonspecial}\,iii).
 	
 	Finally, since $\operatorname{supp}(L)\cap\operatorname{supp}(D)=\emptyset$ and $\gcd(G+L,H-L)$ is non-special, Theorem~\ref{prop_perturbation_hull} applies, and we get
 	\begin{align*}
 		\dim_{\mathbb{F}_q}\operatorname{Hull}\big(\cC(D,G+L)\big)
 		&=\dim_{\mathbb{F}_q}\operatorname{Hull}\big(\cC(D,G)\big)+\frac{1}{2}\big(\deg(|G-H|)-\deg(|G-H+2L|)\big)\\
 		&=\dim_{\mathbb{F}_q}\operatorname{Hull}\big(\cC(D,G)\big)+\deg(\gcd(G+L,H-L))-\deg(\gcd(G,H))\\
 		&=\dim_{\mathbb{F}_q}\operatorname{Hull}\big(\cC(D,G)\big)+\deg(|L| ),
 	\end{align*}
 	where $$\deg(|L|)=\sum_{P\in T_{+}}t_P\deg(P)+\sum_{P\in T_{-}}s_P\deg(P).$$
	This concludes the first part of the theorem. 
 
 	We now prove the second part of the theorem. By Proposition~\ref{thm:upper-bound-hull},
 	$$
 	\dim_{\fq}\hull (\cC(D,G))\leq \frac{1}{2}(n-\deg (R_W)), \quad \text{where}\quad 
 	R_W:=\sum_{\substack{P\in\cP_\cF\setminus\operatorname{supp}(D)\\ v_P(W)\ \text{odd}}} P.$$
 	Now take, in the first part of the proof, the maximal values of $t_P$ and $s_P$ at each place $P$, that is,
 	\[
 	t_P=\left\lfloor\frac{v_P(H)-v_P(G)}{2}\right\rfloor  
 	\qquad\text{and}\qquad
 	s_P=\left\lfloor\frac{v_P(G)-v_P(H)}{2}\right\rfloor.
 	\]
 	This yields
 	\[
 	|v_P(G-H+2L)|=
 	\begin{cases}
 		1, & \text{if } v_P(G-H) \text{ is odd},\\
 		0, & \text{if } v_P(G-H) \text{ is even}
 	\end{cases}
 	\]
 	and therefore $|v_P(G-H+2L)|\leq 1$ for every $P\in \cP_\cF$. Thus, by Proposition~\ref{thm:upper-bound-hull} applied to the code $\cC(D,G+L)$ (see Remark~\ref{rem:upperbound}\,(1)), we obtain
 	$$
 	\dim_{\fq}\hull (\cC(D,G+L))=\frac{1}{2}(n-\deg (R_W)).
 	$$
 \end{proof}

As a consequence of the previous theorem, the bound \eqref{eq_bound_RW} is sharp and the inequality \eqref{eq_bound_max_L} is in fact an equality. Moreover, in Theorem~\ref{hull+} the integers $t_P$ and $s_P$ range independently over $\{0,\dots, m_P \}$ and $\{0,\dots, \mu_P \}$, respectively, where $m_P=\floor*{\frac{v_P(H)-v_P(G)}{2}}$ and $\mu_P=\floor*{\frac{v_P(G)-v_P(H)}{2}}$. Hence, if all the places of $T_+\cup T_{-}=\supp (G-H)$ are rational, every integer of the interval
$$
\left[\dim_{\fq}\hull \cC(D,G),\ \dim_{\fq}\hull \cC(D,G)+\sum_{P\in T_+}m_P+\sum_{P\in T_-}\mu_P\right]
$$
is attained; that is, the hull can be raised \emph{one unit at a time}. This is the mechanism that Section~\ref{sec:kummer} exploits for Kummer extensions.

The previous theorem raises the hull; the next one shows that a similar mechanism lowers it.

\begin{theorem}[Lowering the hull dimension]\label{hull-}
	Let $D$, $G$, $H$ be divisors as in Assumption~\ref{ass:main}, and let $E$ be a non-special divisor such that $E\le\gcd(G,H)$ and $\supp(E)\cap\supp(D)=\emptyset$. Define
	$$
	T_{+}=\supp\big((H-G)_{+}\big)
	\quad\text{and}\quad
	S_E=\supp\big(\gcd(G,H)-E\big).
	$$
	For $P\in S_E$, let
	$
	0\le t_P\le v_P\big(\gcd(G,H)-E\big).
	$
	Set
	$$
	L=\sum_{P\in S_E\setminus T_{+}}t_P\,P-\sum_{P\in S_E\cap T_{+}}t_P\,P .
	$$
	Then $\supp(L)\cap\supp(D)=\emptyset$, the divisor $\gcd(G+L,H-L)$ is non-special, and
	 $$
	\dim_{\fq}\hull\big(\cC(D,G+L)\big)=\dim_{\fq}\hull\big(\cC(D,G)\big)-\sum_{P\in S_E}t_P\deg(P).
	$$
	In particular, taking $t_P=v_P\big(\gcd(G,H)-E\big)$ for every $P\in S_E$, we have
	$$
	\dim_{\fq}\hull\big(\cC(D,G+L)\big)=\deg( E)+1-g.
	$$
\end{theorem}

\begin{proof}
	First, note that for $P_i\in\supp(D)$ we have $v_{P_i}(G)=v_{P_i}(H)=0$, so $v_{P_i}(\gcd(G,H))=0$, and $v_{P_i}(E)=0$ since $\supp(E)\cap\supp(D)=\emptyset$. Hence $P_i\notin S_E$, and so $\supp(L)\cap\supp(D)=\emptyset$.
	
	On the other hand, let $P\in\cP_\cF$. We consider three cases.
	\begin{itemize}
		\item If $P\notin S_E$, then $v_P(L)=0$, so
		\[
		\min\{v_P(G+L),v_P(H-L)\}=\min\{v_P(G),v_P(H)\}.
		\]
		\item If $P\in S_E\setminus T_{+}$, then $v_P(G)\ge v_P(H)$, so $v_P(G)+t_P\ge v_P(H)-t_P$. Hence
		\[
		\min\{v_P(G)+t_P,\ v_P(H)-t_P\}=v_P(H)-t_P=\min\{v_P(G),v_P(H)\}-t_P .
		\]
		\item If $P\in S_E\cap T_{+}$, then $v_P(G)<v_P(H)$, so $v_P(G)-t_P<v_P(H)+t_P$. Hence
		\[
		\min\{v_P(G)-t_P,\ v_P(H)+t_P\}=v_P(G)-t_P=\min\{v_P(G),v_P(H)\}-t_P .
		\]
	\end{itemize}
	Therefore
	\[
	\gcd(G+L,H-L)=\gcd(G,H)-\sum_{P\in S_E}t_P\,P=\gcd(G,H)-|L| .
	\]
	Since $0\le t_P\le v_P(\gcd(G,H)-E)$ for every $P\in S_E$, we have $|L|\le\gcd(G,H)-E$, that is, $E\le\gcd(G+L,H-L)$. Because $E$ is non-special, $\gcd(G+L,H-L)$ is non-special by Lemma~\ref{lem:nonspecial}\,iii).
	
	Finally, since $\supp(L)\cap\supp(D)=\emptyset$ and $\gcd(G+L,H-L)$ is non-special, Theorem~\ref{prop_perturbation_hull} applies. Since $(G+L)+(H-L)=G+H$, we have
	\begin{align*}
		\dim_{\fq}\hull\big(\cC(D,G+L)\big)
		&=\dim_{\fq}\hull\big(\cC(D,G)\big)+\frac{1}{2}\big(\deg(|G-H|)-\deg(|G-H+2L|)\big)\\
		&=\dim_{\fq}\hull\big(\cC(D,G)\big)+\deg(\gcd(G+L,H-L))-\deg(\gcd(G,H))\\
		&=\dim_{\fq}\hull\big(\cC(D,G)\big)-\deg(|L|) ,
	\end{align*}
	where $$\deg(|L|)=\sum_{P\in S_E}t_P\deg(P).$$
	This concludes the first part of the theorem.
	
	For the second part, with the maximal choices $t_P=v_P(\gcd(G,H)-E)$ for every $P\in S_E$, we get $|L|=\gcd(G,H)-E$, and hence $\gcd(G+L,H-L)=E$. Since $E$ is non-special and $\deg (E)\le\deg(\gcd(G,H))<n$, Theorem~\ref{hull nonspecial} gives
	$$
	\dim_{\fq}\hull\big(\cC(D,G+L)\big)=\ell(E)=\deg (E)+1-g .
	$$
\end{proof}

In Theorem~\ref{hull-}, the integers $t_P$ range independently over $\{0,\dots,v_P(\gcd(G,H)-E)\}$ for $P\in S_E$. Since
$$
\sum_{P\in S_E}v_P\big(\gcd(G,H)-E\big)\deg(P)=\dim_{\fq}\hull\big(\cC(D,G)\big)-\big(\deg (E)+1-g\big),
$$
if all the places of $S_E=\supp(\gcd(G,H)-E)$ are rational, then every integer of the interval
$$
\left[\deg( E)+1-g,\ \dim_{\fq}\hull\big(\cC(D,G)\big)\right]
$$
is attained; that is, the hull can be lowered \emph{one unit at a time} down to $\deg (E)+1-g$, and down to $0$ when $\deg (E)=g-1$. This mechanism is used in Section~\ref{sec:kummer} to descend from codes with large hull to LCD codes.

Theorems~\ref{hull+} and~\ref{hull-} are the two faces of one phenomenon. Raising the coefficient of $G$ at a place $P$ where $v_P(G)<v_P(H)$ (or lowering it at a place where $v_P(G)>v_P(H)$) brings $G$ closer to $H$ and increases the common part $\gcd(G,H)$, hence the hull; raising it at a place $P$ where $v_P(G)\ge v_P(H)$ (or lowering it at a place where $v_P(G)<v_P(H)$, as for $P\in S_E\cap T_+$ in Theorem~\ref{hull-}) pushes $G$ further away from $H$ and lowers $\gcd(G,H)$, hence the hull. In both cases, the variation equals the degree of the change of the $\gcd$.

\section{Raising and lowering the hull dimension in Kummer extensions}\label{sec:kummer}

We begin by fixing notation for Kummer extensions; we refer to \cite[Section~3.7]{S2009} and \cite{CMQ2016} for details. Let $m \geq 2$ be a divisor of $q-1$. Let
\[
f(x) = \prod_{i=1}^{r} (x - \alpha_i) \in \fq[x]
\]
be a separable polynomial of degree $r \geq 2$ with $\alpha_1, \dots, \alpha_r \in \fq$ pairwise distinct and $\gcd(m, r) = 1$. Consider the curve $\cX$ over $\fq$ given by the affine equation
\begin{equation}\label{equationX}
\cX \colon \quad y^m = f(x),
\end{equation}
and let $\fq(\cX)$ be its function field. Then $\fq(\cX)/\fq(x)$ is a
\emph{Kummer extension} of degree $m$, and $\fq(\cX)$ has genus
$
g = (m-1)(r-1)/2 \geq 1.
$
Moreover, the place at infinity and the zeros of $x - \alpha_i$ in $\fq(x)$ are totally
ramified in $\fq(\cX)/\fq(x)$. We denote by $Q_\infty$ the unique place of $\fq(\cX)$ lying over the
pole of $x$ (which is also the unique pole of $y$), and by $Q_i$ the unique place of $\fq(\cX)$ lying over the zero of $x - \alpha_i$, for $i = 1, \dots, r$. All of these places are rational. Since these places are totally ramified, we have
\begin{equation}\label{eq:kummerdivisors}
(x-\alpha_i)=m\,Q_i-m\,Q_\infty\quad(1\le i\le r)\qquad\text{and}\qquad (y)=\sum_{i=1}^rQ_i-r\,Q_\infty .
\end{equation}
In Theorem~\ref{improvinghull} and in the rest of this section, this setting is used over the field $\fqs$ and with $m=q+1$, which divides $q^2-1$.

Let $p = \operatorname{char}(\fq)$. A polynomial $f(x) \in \fq[x]$ is called a
\emph{linearized polynomial} if it is of the form $f(x)=\sum_{j=0}^{s}a_jx^{p^j}$ with $a_j\in\fq$. Such a polynomial induces an $\mathbb{F}_p$-linear map on $\overline{\mathbb{F}}_q$. Note that $f$ is separable if and only if $a_0\neq0$.

In what follows, we illustrate an application of
Theorem~\ref{hull+} using the construction of LCD codes given in
\cite{CMQ2025}. More precisely, starting from an AG code
$\cC(D,G)$ with trivial hull (that is, an LCD code), we show
how the hull dimension can be increased by adding a suitable divisor to $G$. The outcome is a family of codes in which the hull dimension can be prescribed at will, up to an explicit maximum.

 \begin{theorem}\label{improvinghull}
Let $\cF/\mathbb{F}_{q^2}(x)$ be a maximal Kummer extension defined by
$$
y^{q+1}=f(x)=\prod_{i=1}^{r}(x-\alpha_i),
$$
where $\alpha_1, \dots, \alpha_r\in\fqs$ are pairwise distinct, $2\le r<q+1$ is coprime with $q+1$, and $f(x)$ is a linearized polynomial. Let $t\ge 2q+2$ be a divisor of $q^2-1$ and define the divisor $D=(y^t-1)_0$.

Then, for each $0\leq \ell\leq r\floor*{\frac{t-1}{2}}+\floor*{\frac{r}{2}}$, there exists a divisor $G_\ell$ such that $\cC(D,G_\ell)$ is an AG code with length $rt$ and
$$
\dim_{\fqs}\hull (\cC(D,G_\ell))=\ell.
$$
More precisely, $G_\ell$ can be chosen of the form $G_\ell=G+\sum_{i=1}^{r-1}\tilde t_iQ_i-\tilde t_rQ_r-\tilde t_{r+1}Q_\infty$, where $G$ is given in \eqref{eq:GHkummer} below, the integers $\tilde t_i$ satisfy the bounds \eqref{eq:boundsti} (with $t_i$ replaced by $\tilde t_i$) and $\sum_{i=1}^{r+1}\tilde t_i=\ell$. For every such choice, $\cC(D,G_\ell)$ is an
$$
\Big[\,rt,\ \ r(q+1)+t-q-1+\delta_\ell,\ \ \ge rt-\deg(G)-\delta_\ell\,\Big]
$$
code, where $\delta_\ell =\sum_{i=1}^{r-1}\tilde t_i-\tilde t_r-\tilde t_{r+1}=\deg(G_\ell)-\deg(G)$.
  \end{theorem}
 \begin{proof} 
 In this case, the function field $\cF$ has genus $g=q(r-1)/2$, and since $r$ is a power of $p$, we have $\gcd(r, q+1)=1$. Since $\cF$ is  maximal over $\fqs$, the divisor $D$ has degree $rt$. Following the notation for Kummer extensions, define the divisors
\begin{equation}\label{eq:GHkummer}
\begin{aligned}
G & = \sum_{i=1}^{r-1} \floor*{ \frac{i(q+1)}{r} } Q_i + (t-q-1) Q_r + (r(q+1)-1)Q_\infty, \quad \mbox{ and }\\
H & =  \sum_{i=1}^{r-1} \left( (t - q- 1) + \floor*{ \frac{(r-i)(q+1)}{r} } \right) Q_i + q Q_r - (q + 1) Q_\infty.
\end{aligned}
\end{equation}
From \cite[Theorem 11]{CMQ2025}, we have that  $\cC(D,G)$ is an $$[ rt, \, r(q+1)+t-q-1  , \, \ge (r-1)(t-q-1)-g+1]$$ LCD AG code with $\cC(D,G)^\perp=\cC(D,H)$. We also have that $\gcd(G,H)$ is a non-special divisor of degree $g-1$. Moreover,
$$
H-G=\sum_{i=1}^{r-1}\left(t-1-2\floor*{\frac{i(q+1)}{r}}\right)Q_i-(t-2q-1)Q_r-(q+r(q+1))Q_\infty,
$$
and since $t\geq 2q+2$, we have
$$
(H-G)_+ =\sum_{i=1}^{r-1}\left(t-1-2\floor*{\frac{i(q+1)}{r}}\right)Q_i\quad \text{and}\quad (G-H)_+=(t-2q-1)Q_r+(q+r(q+1))Q_\infty.$$ 

This yields 
$$T_+=\supp((H-G)_+)=\{Q_1, \dots, Q_{r-1}\}\quad \text{and}\quad  T_-=\supp((G-H)_+)=\{Q_r, Q_\infty\}.$$
Let $t_1,\dots,t_{r+1}\in\N_0$ be such that
\begin{equation}\label{eq:boundsti}
\begin{aligned}
&0\leq t_i\leq m_i:=\floor*{\frac{v_{Q_i}(H)-v_{Q_i}(G)}{2}} =  \floor*{\frac{t-1}{2}}-\floor*{\frac{i(q+1)}{r}}\quad (1\le i\le r-1),\\
&0\leq t_r \leq m_r:=\floor*{\frac{v_{Q_r}(G)-v_{Q_r}(H)}{2}} =  \floor*{\frac{t-2q-1}{2}}= \floor*{\frac{t-1}{2}}-q,\\
&0\leq t_{r+1} \leq m_{r+1}:=\floor*{\frac{v_{Q_\infty}(G)-v_{Q_\infty}(H)}{2}} =  \floor*{\frac{q+r(q+1)}{2}}.
\end{aligned}
\end{equation}

Applying Theorem~\ref{hull+}, we obtain
\begin{equation*}
\dim_{\fqs}\hull \cC(D, G+\sum_{i=1}^{r-1}t_iQ_i-t_r Q_r- t_{r+1} Q_\infty)=\sum_{i=1}^{r+1}t_i.
\end{equation*}

In particular, to obtain the largest possible improvement on the hull, we can choose $t_i= \floor*{\frac{t-1}{2}}-\floor*{\frac{i(q+1)}{r}}$ for every $1\leq i \leq r-1$, $t_r=\floor*{\frac{t-1}{2}}-q$, and $t_{r+1}=\floor*{\frac{q+r(q+1)}{2}}$. This gives

\begin{align*}
\MoveEqLeft \dim_{\fqs}\hull \cC\Big(D,\, G+\textstyle\sum_{i=1}^{r-1}t_iQ_i-t_rQ_r-t_{r+1}Q_\infty\Big)\\
&=\sum_{i=1}^{r-1}\left(\floor*{\frac{t-1}{2}}-\floor*{\frac{i(q+1)}{r}}\right)
  +\left(\floor*{\frac{t-1}{2}}-q\right)
  +\floor*{\frac{q+r(q+1)}{2}}\\
&=\left((r-1)\floor*{\frac{t-1}{2}}-\frac{q(r-1)}{2}\right) +\left(\floor*{\frac{t-1}{2}}-q\right)+\left(\frac{q(r+1)}{2}+\floor*{\frac{r}{2}}\right)\\
&=r\floor*{\frac{t-1}{2}}
  +\left(-\frac{q(r-1)}{2}-q+\frac{q(r+1)}{2}\right)
  +\floor*{\frac{r}{2}}\\
&=r\floor*{\frac{t-1}{2}}+\floor*{\frac{r}{2}},
\end{align*}
where we used $\sum_{i=1}^{r-1}\floor*{\frac{i(q+1)}{r}}=\frac{q(r-1)}{2}$, since $\gcd(r,q+1)=1$.

Let $0\leq \ell\leq r\floor*{\frac{t-1}{2}}+\floor*{\frac{r}{2}}$. It follows from the above that
there exist non-negative integers $\tilde t_1,\dots,\tilde t_{r+1}$ with
$\ell=\tilde t_1+\dots+\tilde t_{r+1}$ such that, setting
\[
G_\ell =G+\sum_{i=1}^{r-1}\tilde t_iQ_i-\tilde t_rQ_r-\tilde t_{r+1}Q_\infty,
\]
the code $\cC(D,G_\ell)$ has hull dimension $\ell$.

It remains to compute the parameters. Put $L_\ell:=G_\ell-G$ and $H_\ell:=H-L_\ell=D-G_\ell+W$. By the proof of Theorem~\ref{hull+}, $\gcd(G_\ell,H_\ell)=\gcd(G,H)+|L_\ell|\ge\gcd(G,H)$, which is non-special; since $G_\ell$ and $H_\ell$ are both $\ge\gcd(G_\ell,H_\ell)$, they are non-special by Lemma~\ref{lem:nonspecial}\, iii). If $\deg(G_\ell)<n$, then $\dim_{\fqs}\cC(D,G_\ell)=\ell(G_\ell)=\deg(G_\ell)+1-g$. Otherwise $\deg(H_\ell)=n+2g-2-\deg(G_\ell)\le 2g-2<n$, and $\dim_{\fqs}\cC(D,G_\ell)=n-\ell(H_\ell)=n-(\deg(H_\ell)+1-g)=\deg(G_\ell)+1-g$. In both cases, since $\deg(G)+1-g=r(q+1)+t-q-1$, we get $\dim_{\fqs}\cC(D,G_\ell)=r(q+1)+t-q-1+\delta_\ell$, and the Goppa bound gives $d\ge rt-\deg(G_\ell)=rt-\deg(G)-\delta_\ell$.
 \end{proof}

The twists used in the proof of Theorem~\ref{improvinghull} are in general not principal, so the codes $\cC(D,G_\ell)$ are in general not equivalent to the LCD code $\cC(D,G)$. However, the principal divisors \eqref{eq:kummerdivisors} are supported precisely on the places $Q_1,\dots,Q_r,Q_\infty$ used in the proof, and they provide \emph{genuine scalings} of $\cC(D,G)$ with large hull. This gives a direct answer to problem~\eqref{problem1} and shows that the bound of Corollary~\ref{cor:principal} is sharp.

\begin{proposition}\label{prop:kummerprincipal}
Keep the notation and hypotheses of Theorem~\ref{improvinghull}, let $G$ be the divisor in \eqref{eq:GHkummer}, and let $m_1,\dots,m_{r+1}$ be as in \eqref{eq:boundsti}. Let $c_1,\dots,c_{r}\in\N_0$ and put $c_\infty=\sum_{i=1}^{r-1}c_i-c_r$. Assume that
$$
c_\infty\ge 0,\qquad (q+1)c_i\le m_i\ \ (1\le i\le r),\qquad (q+1)c_\infty\le m_{r+1},
$$
and set
$$
z=\frac{(x-\alpha_r)^{c_r}}{\prod_{i=1}^{r-1}(x-\alpha_i)^{c_i}}\in\cF^\ast .
$$
Then $z(P)\ne 0,\infty$ for every $P\in\supp(D)$, the code ${\bf x}_z\cdot\cC(D,G)=\cC(D,G-(z))$ is equivalent to the LCD code $\cC(D,G)$, and
$$
\dim_{\fqs}\hull\big({\bf x}_z\cdot\cC(D,G)\big)=2(q+1)\sum_{i=1}^{r-1}c_i .
$$
In particular, if $c_r=0$ and $c_1+\cdots+c_{r-1}\geq 1$, then $[\cF:\fqs(z)]=(q+1)\sum_{i=1}^{r-1}c_i$, so that the variation of the hull dimension under the scaling by $z$ equals $2[\cF:\fqs(z)]$, and the bound of Corollary~\ref{cor:principal} is attained.
\end{proposition}

\begin{proof}
By \eqref{eq:kummerdivisors} with $m=q+1$,
\begin{align*}
(z)&=-\sum_{i=1}^{r-1}c_i\big((q+1)Q_i-(q+1)Q_\infty\big)+c_r\big((q+1)Q_r-(q+1)Q_\infty\big)\\
&=-\sum_{i=1}^{r-1}(q+1)c_iQ_i+(q+1)c_rQ_r+(q+1)c_\infty Q_\infty .
\end{align*}
Hence $z$ has neither zeros nor poles at the places of $\supp(D)$, and Proposition~\ref{prop2210}\,iii) gives ${\bf x}_z\cdot\cC(D,G)=\cC(D,G-(z))$. Let $L:=-(z)$.
This is a divisor of the form considered in the proof of Theorem~\ref{improvinghull}, with $t_i=(q+1)c_i$ for $1\le i\le r$ and $t_{r+1}=(q+1)c_\infty$, and these integers satisfy \eqref{eq:boundsti} by hypothesis. Hence Theorem~\ref{hull+} applies (in particular $z\in U_D(G)$), and since $\dim_{\fqs}\hull\cC(D,G)=0$,
$$
\dim_{\fqs}\hull\cC(D,G-(z))=\sum_{i=1}^{r+1}t_i=(q+1)\Big(\sum_{i=1}^{r-1}c_i+c_r+c_\infty\Big)=2(q+1)\sum_{i=1}^{r-1}c_i .
$$
If $c_r=0$ and $\sum_{i=1}^{r-1}c_i \geq 1$, then $(z)_\infty=(q+1)\sum_{i=1}^{r-1}c_iQ_i$, so $[\cF:\fqs(z)]=\deg((z)_\infty)=(q+1)\sum_{i=1}^{r-1}c_i$.
\end{proof}

The conditions of Proposition~\ref{prop:kummerprincipal} are easy to satisfy. For instance, $c_1=1$ and $c_i=0$ for $i\ge2$ (that is, $z=(x-\alpha_1)^{-1}$) is admissible as soon as $\floor*{\frac{t-1}{2}}-\floor*{\frac{q+1}{r}}\ge q+1$, since $m_{r+1}=\floor*{\frac{q+r(q+1)}{2}}\ge q+1$ always holds for $r\ge2$. In that case the LCD code $\cC(D,G)$ is equivalent to a code with hull dimension $2(q+1)$.

For the Hermitian function field ($r=q$, $t=q^2-1$), we have $m_i=\floor*{\frac{q^2-2}{2}}-i$ for $1\le i\le q-1$, $m_q=\floor*{\frac{q^2-2}{2}}-q$ and $m_{q+1}=\floor*{\frac{q^2+2q}{2}}$. Thus, within the equivalence class of the Hermitian LCD code $\cC(D,G)$, Proposition~\ref{prop:kummerprincipal} produces codes whose hull dimensions are the multiples $2(q+1)s$ for all $s$ in an explicit range, for instance for $0\le s\le \floor*{\frac{q^2+2q}{2(q+1)}}$ when $q\geq4$ by taking the $c_i\in\{0,1\}$ with $c_r=0$.

The codes of Proposition~\ref{prop:kummerprincipal} are among the codes $\cC(D,G'_s)$ of Corollary~\ref{cor:kummerconst} below (with $s=(q+1)\sum_{i=1}^{r-1}c_i$); the point is that, for these particular twists, the codes do not merely have the same parameters as $\cC(D,G)$, but are genuinely \emph{equivalent} to it.

\begin{remark}\label{rem:hermitianmax}
	Applying Theorem~\ref{improvinghull} to the Hermitian function field defined by $y^{q+1}=x^q+x$ over $\fqs$, with $q \geq 3$ (see \cite[Example~6.3.6]{S2009}, \cite{SH1988} and \cite{GSX2000}), and choosing $t=q^2-1$, we obtain AG codes of length $n=q^3-q$. In particular, for the divisor $G=\sum_{i=1}^{q-1}iQ_i+(q^2-q-2)Q_q+(q^2+q-1)Q_\infty$ used in the proof of Theorem~\ref{improvinghull}, we obtain
	\begin{equation*}
		\max_{\substack{L\in\mathrm{Div}(\cF)\\ \supp(L)\cap\supp(D)=\emptyset}}\dim_{\fqs}\hull\big(\cC(D,G+L)\big)\geq
		\begin{cases}
			\dfrac{q^3-2q-1}{2}, & q \text{ odd},\\[8pt]
			\dfrac{q^3-q}{2}, & q \text{ even}.
		\end{cases}
	\end{equation*}
	This value is attained by the code $\cC(D,G+L)$, where the maximal choices in the proof of Theorem~\ref{improvinghull} give
	\begin{equation*}
		G+L=
		\begin{cases}
			\dfrac{q^2-3}{2}\,(Q_1+\cdots+Q_{q-1})+\dfrac{q^2-1}{2}\,(Q_q+Q_\infty), & q \text{ odd},\\[10pt]
			\dfrac{q^2-2}{2}\,(Q_\infty+Q_1+\cdots+Q_q), & q \text{ even}.
		\end{cases}
	\end{equation*}
	Note that, for every divisor $L$ such that $\supp(L)\cap\supp(D)=\emptyset$, we have
	$$
	\dim_{\fqs}\hull\big(\cC(D,G+L)\big)\leq\frac{n}{2}=\frac{q^3-q}{2}.
	$$
	Therefore, equality holds for $q$ even, and in this case $\cC(D,G+L)$ is a self-dual code. For $q$ odd, the canonical divisor used here is $W=-D+(q^2-2)(Q_\infty+Q_1+\cdots+Q_q)$ (see the proof of Theorem~\ref{thm:hermitian} below), and $q^2-2$ is odd, so that $R_W=Q_\infty+Q_1+\cdots+Q_q$ and $\frac12(n-\deg(R_W))=\frac{q^3-2q-1}{2}$. Hence, by \eqref{eq_bound_max_L} and Theorem~\ref{hull+}, for $q$ odd the first inequality of this remark is an \emph{equality} when the maximum is taken over the set $\mathcal{T}_D(G)$ of Remark~\ref{rem:upperbound}.
	\end{remark}

Theorem~\ref{improvinghull} starts from an LCD code and raises the hull. For the Hermitian function field with $q$ even and $t=q^2-1$, Remark~\ref{rem:hermitianmax} shows that the largest possible hull dimension is attained by a self-dual code. We now go in the opposite direction. Starting from this self-dual code, we lower the hull dimension one unit at a time until we reach an LCD code. The end of this descent is governed by a non-special divisor $E$ of degree $g-1$ supported on the places $Q_\infty,Q_1,\dots,Q_q$. Such divisors are completely characterised in \cite[Corollary~4]{MNQ2026}; in particular, they always exist. Different choices of $E$, and of the signs of the twist, lead to LCD codes of different dimensions; one of them is the LCD code of the proof of Theorem~\ref{improvinghull}.

\begin{theorem}\label{thm:hermitian}
	Let $q$ be an even prime power, and let $\cH=\fqs(x,y)$ be the Hermitian function field defined by $y^{q+1}=x^q+x$. Set
	$$
	D=(y^{q^2-1}-1)_0
	\qquad\text{and}\qquad
	G=\frac{q^2-2}{2}\,(Q_\infty+Q_1+\cdots+Q_q).
	$$
	Then $\cC(D,G)$ is a self-dual AG code of length $n=q^3-q$. Let $E$ be a non-special divisor of degree $g-1$ with $\supp(E)\subseteq\{Q_\infty,Q_1,\dots,Q_q\}$ and $E\leq G$, and let $L$ be a divisor with $\supp(L)\subseteq\{Q_\infty,Q_1,\dots,Q_q\}$ and $|L|\leq G-E$. Then:
	\begin{enumerate}[i)]
		\item $\dim_{\fqs}\hull\big(\cC(D,G+L)\big)=\dfrac{n}{2}-\deg(|L|)$,
		\item $\cC(D,G+L)$ is an $\Big[\,n,\ \dfrac{n}{2}+\deg( L),\ d\geq\dfrac{n}{2}-g+1-\deg( L)\,\Big]$ code, and
		\item $\cC(D,G+L)$ is an LCD code if and only if $|L|=G-E$.
	\end{enumerate}
	In particular, the hull dimension can be lowered one unit at a time from $n/2$ down to $0$.
\end{theorem}

\begin{proof}
	The Hermitian function field $\cH$ has genus $g=q(q-1)/2$ and the divisor $D$ is the sum of all rational places of $\cH$ other than $Q_\infty,Q_1,\dots,Q_q$, so $n=\deg (D)=q^3+1-(q+1)=q^3-q$.

	The canonical divisor can be found in the proof of \cite[Theorem~11]{CMQ2025}; for the reader's convenience, we compute it here (the computation is valid for every $q$, including $q=2$). The function $u=y^{q^2-1}-1$ has a unique pole at $Q_\infty$ of order $n$, and its $n$ zeros, which are the places of $D$, are all simple. The differential $\eta=\frac{du}{u}$ satisfies $v_P(\eta)=-1$ and $\eta_P(1)=\operatorname{res}_P(du/u)=v_P(u)=1$ for every $P\in\supp(D)$. The different of $\cH/\fqs(x)$ is given by $q\sum_{i=1}^qQ_i+qQ_\infty$, so that $(dx)=q\sum_{i=1}^qQ_i-(q+2)Q_\infty$. Differentiating $y^{q+1}=x^q+x$ gives $y^q\,dy=dx$, hence $(dy)=(q^2-q-2)Q_\infty=(2g-2)Q_\infty$. Since $du=(q^2-1)y^{q^2-2}dy$, we obtain
	$$
	W:=(\eta)=(y^{q^2-2}dy)-(u)=-D+(q^2-2)(Q_\infty+Q_1+\cdots+Q_q).
	$$
	Hence $D+W=2G$, that is, $H=D-G+W=G$, so $\cC(D,G)^\perp=\cC(D,G)$, that is, $\cC(D,G)$ is self-dual. Moreover, $\deg (G)=\frac{n}{2}+g-1>2g-2$, so $\gcd(G,H)=G$ is non-special, and $D$, $G$, $H$ and $W$ satisfy Assumption~\ref{ass:main}.
	
	Note that $G-E\geq 0$ and $\deg (G-E)=\frac{n}{2}$. Let $L$ be as in the statement. Since $G=H$, we have
	$$
	\gcd(G+L,H-L)=\gcd(G+L,G-L)=G-|L|\ \geq E,
	$$
	so $\gcd(G+L,H-L)$ is non-special by Lemma~\ref{lem:nonspecial}\,iii). Moreover, $\supp(L)\cap\supp(D)=\emptyset$, and $\supp(L)\cap\supp(G-H)=\emptyset$ because $G-H=0$. Hence Theorem~\ref{prop_perturbation_hull}\,b) gives
	$$
	\dim_{\fqs}\hull\big(\cC(D,G+L)\big)=\dim_{\fqs}\hull\big(\cC(D,G)\big)-\deg(|L|)=\frac{n}{2}-\deg(|L|),
	$$
	which proves i).
	
	For ii), put $G'=G+L$ and $H'=G-L$. Both divisors are greater than or equal to the non-special divisor $G-|L|$, so both are non-special by Lemma~\ref{lem:nonspecial}\,iii). Since $H=G$, we also have $H'=H-L=D-G'+W$, so $\cC(D, G')^\perp=\cC(D, H')$. If $\deg (G')<n$, then $\dim_{\fqs}\cC(D,G')=\ell(G')=\deg (G')+1-g$. Otherwise $\deg( H')=n+2g-2-\deg( G')\leq 2g-2<n$, and
	$$
	\dim_{\fqs}\cC(D,G')=n-\dim_{\fqs}\cC(D,H')=n-\big(\deg (H')+1-g\big)=\deg (G')+1-g.
	$$
	In both cases $\dim_{\fqs}\cC(D,G+L)=\deg( G)+\deg( L)+1-g=\frac{n}{2}+\deg( L)$. The bound on the minimum distance is the Goppa bound $d\geq n-\deg(G+L)$.
	
	For iii), by i), the code is LCD if and only if $\deg(|L|)=\frac{n}{2}=\deg (G-E)$. Since $0\leq |L|\leq G-E$, this holds if and only if $|L|=G-E$.
	
	Finally, since $Q_\infty,Q_1,\dots,Q_q$ are rational places and $\deg (G-E)=\frac{n}{2}$, for every $0\leq j\leq\frac{n}{2}$ there is a divisor $L$ with $\supp(L)\subseteq\{Q_\infty,Q_1,\dots,Q_q\}$, $|L|\leq G-E$ and $\deg(|L|)=j$.
\end{proof}

We now illustrate how to descend from the codes of maximal hull dimension of Remark~\ref{rem:hermitianmax} down to LCD codes. In both examples, the descent is governed by the same family of non-special divisors $E_s$ of degree $g-1$, supported on $Q_\infty,Q_1,\dots,Q_q$ and defined below, and for $s=0$ we recover the LCD code in the proof of Theorem~\ref{improvinghull}. For $q$ even, the code of maximal hull dimension is self-dual, and we apply Theorem~\ref{thm:hermitian}. For $q$ odd, this code is not self-dual, so Theorem~\ref{thm:hermitian} is no longer available, and we use Theorem~\ref{hull-} instead.

\begin{example}\label{ex:hermitianLCD}
	Let $q\geq 4$ be even, and let
	$$
	D=(y^{q^2-1}-1)_0
	\qquad\text{and}\qquad
	G=\frac{q^2-2}{2}\,(Q_\infty+Q_1+\cdots+Q_q)
	$$
	be the divisors of Theorem~\ref{thm:hermitian}. In particular, $n=q^3-q$, $g=q(q-1)/2$, and $\cC(D,G)$ is self-dual.
	
	For every integer $s$ with $-\frac{q}{2}\leq s\leq\frac{q}{2}-1$, consider the divisor
	$$
	E_s=\big(1+(q+1)s\big)Q_1+\sum_{i=2}^{q}iQ_i-(q+1)(1+s)Q_\infty .
	$$
	Then $E_s$ is a non-special divisor of degree $g-1$ by \cite[Corollary~4]{MNQ2026}. We also have $E_s\leq G$.
	
	Let
	$$
	L_s=-\Big(\frac{q^2-4}{2}-(q+1)s\Big)Q_1-\sum_{i=2}^{q-1}\Big(\frac{q^2-2}{2}-i\Big)Q_i+\frac{q^2-2q-2}{2}\,Q_q+\Big(\frac{q^2+2q}{2}+(q+1)s\Big)Q_\infty .
	$$
	A direct computation shows $|L_s|=G-E_s$. Let
	$$
	G_s:=G+L_s=\big(1+(q+1)s\big)Q_1+\sum_{i=2}^{q-1}iQ_i+(q^2-q-2)Q_q+\big(q^2+q-1+(q+1)s\big)Q_\infty .
	$$
	By Theorem~\ref{thm:hermitian}\,ii) and iii), $\cC(D,G_s)$ is an LCD code with parameters
	$$
	\Big[\,q^3-q,\ \ 2q^2-2+2(q+1)s,\ \ \geq q^3-\frac{5q^2+q}{2}+3-2(q+1)s\,\Big].
	$$
	For $s=0$ we obtain
	$$
	E_0=\sum_{i=1}^{q}iQ_i-(q+1)Q_\infty \quad \text{and}\quad G_0=\sum_{i=1}^{q-1}iQ_i+(q^2-q-2)Q_q+(q^2+q-1)Q_\infty.
	$$
	In this case, we recover the LCD code of the proof of Theorem~\ref{improvinghull} for the Hermitian function field, that is, with $r=q$ and $t=q^2-1$. For $s\neq 0$ we obtain LCD codes of different dimensions. Since $\gcd(G_s,D-G_s+W)=E_s$ is a non-special divisor of degree $g-1$, these codes fit into the general framework of \cite{CLM2024,MNQ2026,CMQ2025}, where LCD and linear complementary pair (LCP) codes are built from such divisors; what the present construction adds is that they are reached by an explicit descent, through all intermediate hull dimensions, from a self-dual code. We point out that, for $s\neq 0$, the parameters of the LCD codes obtained in this example differ from those in~\cite{CMQ2025} and~\cite{CLM2024}.
\end{example}

	\begin{example}\label{ex:hermitianLCDodd}
		Let $q\geq 3$ be odd, and let $\cH$ be the Hermitian function field, $Q_\infty,Q_1,\dots,Q_q$ and $D=(y^{q^2-1}-1)_0$ be defined as in Theorem~\ref{thm:hermitian}, so that $n=q^3-q$ and $g=q(q-1)/2$. By Remark~\ref{rem:hermitianmax}, the code $\cC(D,G)$, where
		$$
		G=\frac{q^2-3}{2}\,(Q_1+\cdots+Q_{q-1})+\frac{q^2-1}{2}\,(Q_q+Q_\infty),
		$$
		has hull dimension $\frac{q^3-2q-1}{2}$. 
		The differential $\eta=\frac{d(y^{q^2-1}-1)}{y^{q^2-1}-1}$ satisfies $v_P(\eta)=-1$ and $\eta_P(1)=1$ for every $P\in\supp(D)$, and
	 $W:=(\eta)=-D+(q^2-2)(Q_\infty+Q_1+\cdots+Q_q)$. Therefore,
		$$
		H=D-G+W=\frac{q^2-1}{2}\,(Q_1+\cdots+Q_{q-1})+\frac{q^2-3}{2}\,(Q_q+Q_\infty).
		$$
		This yields
		$$
		\gcd(G,H)=\frac{q^2-3}{2}\,(Q_\infty+Q_1+\cdots+Q_q).
		$$
		
		For every integer $s$ with $-\frac{q-1}{2}\leq s\leq\frac{q-3}{2}$, let $E_s$ be the divisor given by
		$$
		E_s=\big(1+(q+1)s\big)Q_1+\sum_{i=2}^{q}iQ_i-(q+1)(1+s)Q_\infty .
		$$
		As in Example~\ref{ex:hermitianLCD}, $E_s$ is a non-special divisor of degree $g-1$ by \cite[Corollary~4]{MNQ2026}. Moreover, $E_s\leq\gcd(G,H)$. 
		Here $T_+=\supp\big((H-G)_+\big)=\{Q_1,\dots,Q_{q-1}\}$. Taking $t_P=v_P\big(\gcd(G,H)-E_s\big)$ for every $P\in S_{E_s}=\supp\big(\gcd(G,H)-E_s\big)$ in Theorem~\ref{hull-}, we obtain
		$$
		L_s=-\Big(\frac{q^2-5}{2}-(q+1)s\Big)Q_1-\sum_{i=2}^{q-1}\Big(\frac{q^2-3}{2}-i\Big)Q_i+\frac{q^2-2q-3}{2}\,Q_q+\Big(\frac{q^2+2q-1}{2}+(q+1)s\Big)Q_\infty ,
		$$
		and therefore
		$$
		G_s:=G+L_s=\big(1+(q+1)s\big)Q_1+\sum_{i=2}^{q-1}iQ_i+(q^2-q-2)Q_q+\big(q^2+q-1+(q+1)s\big)Q_\infty,
		$$
		which coincides with the divisor $G_s$ of Example~\ref{ex:hermitianLCD}.

		By Theorem~\ref{hull-}, the hull dimension of $\cC(D,G_s)$ is $\deg (E_s)+1-g=0$, so $\cC(D,G_s)$ is an LCD code. Moreover, $G_s$ and $D-G_s+W$ are both greater than or equal to the non-special divisor $E_s$. Hence, as in the proof of Theorem~\ref{thm:hermitian}\, ii), $\dim_{\fqs}\cC(D,G_s)=\deg (G_s)+1-g$. Therefore $\cC(D,G_s)$ is an LCD code with parameters
		$$
		\Big[\,q^3-q,\ \ 2q^2-2+2(q+1)s,\ \ \geq q^3-\frac{5q^2+q}{2}+3-2(q+1)s\,\Big].
		$$
		For $s=0$, we again recover the LCD code in the proof of Theorem~\ref{improvinghull} with $r=q$ and $t=q^2-1$, now for $q$ odd.
	\end{example}

We close this section by returning to the Kummer extensions of Theorem~\ref{improvinghull}, and we show that there the hull can be tuned \emph{without changing the length, dimension and designed distance of the code}: it suffices to balance the raising twists at the places $Q_1,\dots,Q_{r-1}$ against those at $Q_r$ and $Q_\infty$, so that the twisting divisor has degree zero.

\begin{corollary}\label{cor:kummerconst}
Keep the notation and hypotheses of Theorem~\ref{improvinghull}, let $G$ be the divisor defined in its proof, and set
$$
A=(r-1)\floor*{\frac{t-1}{2}}-\frac{q(r-1)}{2}
\qquad\text{and}\qquad
B=\floor*{\frac{t-1}{2}}-q+\floor*{\frac{q+r(q+1)}{2}}.
$$
Then, for every integer $s$ with $0\le s\le\min\{A,B\}$, there exists a divisor $G'_s$ with $\supp(G'_s)\cap\supp(D)=\emptyset$ and $\deg(G'_s)=\deg(G)$ such that $\cC(D,G'_s)$ is an
$$
\big[\,rt,\ r(q+1)+t-q-1\,\big]
$$
code with designed distance $rt-\deg(G)$, that is, with the same length, dimension and designed distance as the LCD code $\cC(D,G)$, and
$$
\dim_{\fqs}\hull\big(\cC(D,G'_s)\big)=2s.
$$
\end{corollary}

\begin{proof}
In the proof of Theorem~\ref{improvinghull}, the integers $t_1,\dots,t_{r+1}$ range independently over the intervals $[0,m_i]$, where $m_i=\floor*{\frac{t-1}{2}}-\floor*{\frac{i(q+1)}{r}}$ for $1\le i\le r-1$, $m_r=\floor*{\frac{t-1}{2}}-q$ and $m_{r+1}=\floor*{\frac{q+r(q+1)}{2}}$. Since $\sum_{i=1}^{r-1}\floor*{\frac{i(q+1)}{r}}=\frac{q(r-1)}{2}$, we have $\sum_{i=1}^{r-1}m_i=A$ and $m_r+m_{r+1}=B$. Let $0\le s\le\min\{A,B\}$. We may then choose $t_1,\dots,t_{r-1}$ with $\sum_{i=1}^{r-1}t_i=s$, and $t_r,t_{r+1}$ with $t_r+t_{r+1}=s$. Set
$$
L=\sum_{i=1}^{r-1}t_iQ_i-t_rQ_r-t_{r+1}Q_\infty
\qquad\text{and}\qquad
G'_s:=G+L.
$$
Since the places $Q_1,\dots,Q_r,Q_\infty$ are rational, we have $\deg(L)=s-s=0$; hence $\deg(G'_s)=\deg(G)$ and $\supp(G'_s)\cap\supp(D)=\emptyset$. As in the proof of Theorem~\ref{improvinghull}, Theorem~\ref{hull+} gives $\dim_{\fqs}\hull(\cC(D,G'_s))=\sum_{i=1}^{r+1}t_i=2s$.

It remains to compare the parameters. Using again $\sum_{i=1}^{r-1}\floor*{\frac{i(q+1)}{r}}=\frac{q(r-1)}{2}=g$, we get
$$
\deg(G)=g+t-q-2+r(q+1).
$$
On the one hand, $g=\frac{q(r-1)}{2}<r(q+1)$, hence $\deg(G)>2g-2$. On the other hand, since $t\ge 2q+2$,
$$
rt-\deg(G)=(r-1)t-\big(g+r(q+1)-q-2\big)\ge(r-1)(2q+2)-\big(g+r(q+1)-q-2\big)=\frac{q(r-1)}{2}+r>0 .
$$
Thus $2g-2<\deg(G'_s)=\deg(G)<rt=n$, and Proposition~\ref{prop:dimcode}\,ii) gives
$$
\dim_{\fqs}\cC(D,G'_s)=\deg(G)+1-g=r(q+1)+t-q-1=\dim_{\fqs}\cC(D,G),
$$
while the Goppa bound gives the designed distance $n-\deg(G'_s)=rt-\deg(G)$ for both codes.
\end{proof}

In other words, within a family of AG codes with \emph{fixed} length, dimension and designed distance, the hull dimension takes every even value between $0$ and $2\min\{A,B\}$. Two comments are in order. First, the codes $\cC(D,G'_s)$ share the same \emph{designed} distance, but their true minimum distances may differ; when $s$ is a multiple of $q+1$ and the twist is chosen as in Proposition~\ref{prop:kummerprincipal}, the codes are even equivalent to $\cC(D,G)$, hence have the same minimum distance. Second, the construction only produces \emph{even} hull dimensions, because twists of the form used in Theorem~\ref{hull+} with $\deg(L)=0$ satisfy $\deg(|L|)=2\sum_{i=1}^{r-1}t_i$. Whether odd hull dimensions can also be reached at fixed parameters, for instance by using places of higher degree, is left open.

\begin{remark}[Entanglement-assisted quantum codes]\label{rem:EAQECC}
Recall that from a linear $[n,k,d]$ code $\cC$ over $\fq$ one obtains an entanglement-assisted quantum error-correcting code (EAQECC) with parameters
$$
[[\,n,\ k-\dim_{\fq}\hull(\cC),\ d;\ n-k-\dim_{\fq}\hull(\cC)\,]]_q,
$$
see \cite{GJG2018,LEGL2023}. The results of this section therefore give explicit EAQECCs with a prescribed number of ebits. For instance:

(1) In the situation of Theorem~\ref{thm:hermitian} ($q$ even, $n=q^3-q$, $g=q(q-1)/2$), let $L$ be a divisor with $\supp(L)\subseteq\{Q_\infty,Q_1,\dots,Q_q\}$ and $|L|\le G-E$, and put $a:=\deg(L)$ and $b:=\deg(|L|)$, so that $|a|\le b$ and $a\equiv b\pmod 2$. Then $\cC(D,G+L)$ has dimension $\frac n2+a$ and hull dimension $\frac n2-b$, which gives EAQECCs with parameters
$$
\Big[\Big[\,q^3-q,\ \ a+b,\ \ \ge\tfrac{n}{2}-g+1-a;\ \ b-a\,\Big]\Big]_{q^2}.
$$
In particular, for an effective $L$ ($a=b$) no ebit is needed, and in general the number of ebits $b-a=2\deg\big((-L)_+\big)$ is twice the degree of the negative part of $L$.

(2) In the situation of Corollary~\ref{cor:kummerconst}, the codes $\cC(D,G'_s)$ give EAQECCs with parameters
$$
\Big[\Big[\,rt,\ \ r(q+1)+t-q-1-2s,\ \ \ge rt-\deg(G);\ \ rt-r(q+1)-t+q+1-2s\,\Big]\Big]_{q^2},\qquad 0\le s\le\min\{A,B\},
$$
that is, at fixed length and fixed classical parameters, the number of ebits can be decreased by any even amount up to $2\min\{A,B\}$.
\end{remark}

\section{Hull variation without the non-speciality condition on the greatest common divisor}\label{sec:general}

All the results of the previous sections rest on the non-speciality of $\gcd(G,H)$, and most of them also on that of $\gcd(G+L,H-L)$, which is what converts the dimension of an intersection of codes into a degree. When one (or both) of these assumptions is dropped, Corollary~\ref{shaskadual} shows that the dimension of the hull acquires the extra summand $\dim((\cL(G)+\cL(H))\cap \cL(W-\gcd(G,H)))$, and an exact formula of the type \eqref{difference} is no longer expected. We therefore aim at \emph{bounds}.

If we relax the condition that $\gcd(G+L,H-L)$ be non-special in
Theorem~\ref{prop_perturbation_hull}, we obtain the following result.
\begin{proposition}\label{prop:general}
	 Let $D,G$ and $H$ be the divisors defined in Assumption \ref{ass:main}. Suppose moreover that $2g-2<\deg(G)<n=\deg(D)$.
	Let $L$ be a divisor such that $\supp(L)\cap\supp(D)=\emptyset$ and $2g-2<\deg(G+L)<n$.
	Then 
	\begin{equation*}
	\begin{split}
		&\dim_{\fq} \hull (\cC(D,G+L))-\dim_{\fq} \hull (\cC(D,G))\\
		&\quad=\frac{1}{2}\big(\deg(|G-H|)-\deg(|G-H+2L|)\big)+\ell\big(W-\gcd(G+L,H-L)\big)\\
		&\quad+\dim_{\fq}\Big(\big(\cL(G+L)+\cL(H-L)\big)\cap \cL\big(W-\gcd(G+L,H-L)\big)\Big).
	\end{split}
	\end{equation*}
\end{proposition}

\begin{proof}The result follows from Corollary~\ref{shaskadual}, applied to $G$ and to $G+L$, together with \eqref{deggcdGH}, \eqref{deggcdGHL} and the Riemann--Roch theorem, which give
	$$
	\ell\big(\gcd(G,H)\big)=\frac{1}{2}\big(n-\deg(|G-H|)\big)+i\big(\gcd(G,H)\big)
	$$
	and
	$$
	\ell\big(\gcd(G+L,H-L)\big)=\frac{1}{2}\big(n-\deg(|G-H+2L|)\big)+\ell\big(W-\gcd(G+L,H-L)\big).
	$$
	Moreover, since $\gcd(G,H)$ is non-special, we have $\ell\big(W-\gcd(G,H)\big) = i\big(\gcd(G,H)\big) = 0$. Combining these facts yields
	\begin{equation*}
		\begin{split}
			&\dim_{\fq} \hull (\cC(D,G+L))-\dim_{\fq} \hull (\cC(D,G))\\
			&\quad=\ell(\gcd(G+L,H-L))
			+\dim_{\fq}\big((\cL(G+L)+\cL(H-L))\cap \cL(W-\gcd(G+L,H-L))\big)\\
			&\qquad-\ell(\gcd(G,H))
			-\dim_{\fq}\big((\cL(G)+\cL(H))\cap \cL(W-\gcd(G,H))\big)\\
			&\quad=\frac{1}{2}\big(\deg(|G-H|)-\deg(|G-H+2L|)\big)+\ell\big(W-\gcd(G+L,H-L)\big)\\
			&\qquad+\dim_{\fq}\Big(\big(\cL(G+L)+\cL(H-L)\big)\cap \cL\big(W-\gcd(G+L,H-L)\big)\Big).
		\end{split}
	\end{equation*}
\end{proof}
Since the last two summands in Proposition~\ref{prop:general} are non-negative, and vanish when $\gcd(G+L,H-L)$ is non-special, the right-hand side of \eqref{difference} is always a \emph{lower bound} for the variation when only $\gcd(G,H)$ is assumed to be non-special:
$$
\dim_{\fq} \hull (\cC(D,G+L))-\dim_{\fq} \hull (\cC(D,G))\ \ge\ \tfrac12\big(\deg(|G-H|)-\deg(|G-H+2L|)\big),
$$
with equality if and only if $\gcd(G+L,H-L)$ is non-special.

Recall that every divisor of degree greater than $2g-2$ is non-special.
Hence the non-speciality of $\gcd(G,H)$ can be replaced by the condition
$\deg(\gcd(G,H))>2g-2$, which leads to the following proposition.

\begin{proposition}\label{prop:uncond}
Let $D,G,H$ be as in Assumption~\ref{ass:main}\,i)--iii) and suppose that $\deg(\gcd(G,H))> 2g-2$. Then $\gcd(G,H)$ is non-special and
$$
\dim_{\fq}\hull \cC(D,G)=\tfrac12\big(n-\deg(|G-H|)\big)\ \ge g .
$$
\end{proposition}
\begin{proof}
A divisor of degree larger than $2g-2$ is non-special (Lemma~\ref{lem:nonspecial}\,i)), so Assumption~\ref{ass:main}\,iv) holds. By Theorem~\ref{hull nonspecial}, $\dim_{\fq}\hull \cC(D,G)=\ell(\gcd(G,H))$, because $$\deg(\gcd(G,H))=\frac12(n+2g-2-\deg(|G-H|))\le\frac12(n+2g-2)<n.$$ The Riemann--Roch theorem gives $\ell(\gcd(G,H))=\deg(\gcd(G,H))+1-g$, and the stated formula follows from $\deg(\gcd(G,H))=\frac12(n+2g-2-\deg(|G-H|))$. Finally, $\deg(\gcd(G,H))\ge 2g-1$ gives $\dim_{\fq} \hull(\cC(D,G))\ge g$.
\end{proof}

We now consider a twist by a divisor $L$ satisfying condition i) of Theorem~\ref{prop_perturbation_hull}, but not necessarily condition ii).

\begin{proposition}\label{prop:onesided}
Let $D$, $G$, $H$ and $W$ be as in Assumption~\ref{ass:main}, and let $L$ be a divisor with $\supp(L)\cap\supp(D)=\emptyset$. Then
$$
\dim_{\fq}\hull\big(\cC(D,G)\big)-\dim_{\fq}\hull\big(\cC(D,G+L)\big)\leq\deg(|L|).
$$
\end{proposition}

\begin{proof}
Since $\supp(L)\cap\supp(D)=\emptyset$, we have $\cC(D,G+L)^\perp=\cC(D,D-(G+L)+W)=\cC(D,H-L)$. Hence
\begin{align*}
	\hull\big(\cC(D,G+L)\big)&=\cC(D,G+L)\cap \cC(D,G+L)^\perp\\
	&=\cC(D,G+L)\cap \cC(D,H-L)\\
	&\supseteq \cC\big(D,\gcd(G+L,H-L)\big).
\end{align*}
For any divisors $A,B$ we have $\gcd(A,B)\leq\frac{1}{2}(A+B)$ coefficientwise. Since $(G+L)+(H-L)=D+W$, this gives
$$
\deg(\gcd(G+L,H-L))\leq\frac{n+2g-2}{2}<n,
$$
so $\dim \cC\big(D,\gcd(G+L,H-L)\big)=\ell\big(\gcd(G+L,H-L)\big)$. Hence
\begin{align*}
	\dim_{\fq}\hull\big(\cC(D,G+L)\big)&\geq\dim \cC\big(D,\gcd(G+L,H-L)\big)\\
	&=\ell\big(\gcd(G+L,H-L)\big)\\
	&\geq\deg (\gcd(G+L,H-L))+1-g\\
	&\geq\deg (\gcd(G,H))-\deg(|L|)+1-g\\
	&=\dim_{\fq}\hull\big(\cC(D,G)\big)-\deg(|L|),
\end{align*}
where the fourth line follows from the place-by-place inequality $\min\{a+l,\,b-l\}\ge\min\{a,b\}-|l|$ for integers $a,b,l$, applied with $a=v_P(G)$, $b=v_P(H)$, $l=v_P(L)$, and the last line from Theorem~\ref{hull nonspecial} and the non-speciality of $\gcd(G,H)$. This proves the claim.
\end{proof}

The roles of $G$ and $G+L$ can also be exchanged in the following case. If $\gcd(G+L,H-L)$ is non-special but $\gcd(G,H)$ is not, Proposition~\ref{prop:onesided} applied to the base divisor $G+L$ and the twist $-L$ gives the complementary estimate
$$
\dim_{\fq}\hull \cC(D,G+L)-\dim_{\fq}\hull \cC(D,G)\le\deg(|L|).
$$
Thus a one-sided bound is available as soon as \emph{one} of the two gcds is non-special, and the two-sided bound of Theorem~\ref{prop_perturbation_hull}\,a) is recovered when both are.

\begin{proposition}\label{prop:sandwich}
Let $D,G,H$ be as in Assumption~\ref{ass:main}\,i)--iii) with $2g-2<\deg (G)<n$. Then
$$
\tfrac12\big(n-\deg(|G-H|)\big)+i(\gcd(G,H)) \le \dim_{\fq}\hull \cC(D,G) \le \tfrac12\big(n-\deg(|G-H|)\big)+2\,i(\gcd(G,H)).
$$
In particular, $\dim_{\fq}\hull \cC(D,G)\ge\frac12\big(n-\deg(|G-H|)\big)$, with equality if and only if $\gcd(G,H)$ is non-special.
\end{proposition}
\begin{proof}
By Corollary~\ref{shaskadual}, $\dim_{\fq}\hull \cC(D,G)=\ell(\gcd(G,H))+\delta$, where $0\le\delta\le\ell(W-\gcd(G,H))=i(\gcd(G,H))$. By the Riemann--Roch theorem, 
$$\ell(\gcd(G,H))=\deg(\gcd(G,H))+1-g+i(\gcd(G,H)).$$ 
By \eqref{deggcdGH}, we have $$\deg(\gcd(G,H))+1-g=\frac12(n-\deg(|G-H|)).$$ 
Hence $\dim_{\fq}\hull \cC(D,G)=\frac12\big(n-\deg(|G-H|)\big)+i(\gcd(G,H))+\delta$ with $0\le\delta\le i(\gcd(G,H))$, which gives both inequalities. Consequently, $\dim_{\fq}\hull \cC(D,G)=\frac12\big(n-\deg(|G-H|)\big)$ if and only if $i(\gcd(G,H))=\delta=0$, that is, if and only if $\gcd(G,H)$ is non-special (in which case $\delta=0$ automatically).
\end{proof}

\section{Perspectives for future work}\label{sec:conclusion-perspectives}
We conclude this work with directions suggested by the present results.
\begin{enumerate}[i)]
\item \emph{Genericity of non-speciality.} 
Note that Proposition~\ref{prop:uncond} makes non-speciality automatic when $\deg(\gcd(G,H))\ge 2g-1$. A quantitative study of how often $\gcd(G,H)$ is non-special, for instance via the existence results of \cite{BL2006}, would indicate how much of the hull spectrum Theorem~\ref{prop_perturbation_hull} captures.
\item \emph{Entanglement-assisted quantum codes.} 
The number of ebits required by an EAQECC built from a code $\cC$ is $c=n-k-\dim\hull(\cC)$, see \cite{GJG2018,LEGL2023}. Theorems~\ref{hull+}, \ref{hull-}, \ref{improvinghull} and~\ref{thm:hermitian} thus produce EAQECCs with a prescribed number of ebits; by Corollary~\ref{cor:kummerconst}, this can even be achieved at fixed classical parameters, see Remark~\ref{rem:EAQECC}. A systematic comparison of the resulting parameters with known tables is a natural next step.
\item \emph{Hermitian hull.} Hermitian hulls of AG codes have been studied, for instance, in the rational case in \cite{SEL2026}. A dynamical analogue of the present theory for the Hermitian inner product over $\mathbb{F}_{q^2}$, in which the role of $G\mapsto G+L$ is played by twists compatible with the $q$-th power Frobenius, remains to be developed.
\item \emph{Complexity.} Combining Sendrier's analysis \cite{S1997} with our control of $\dim\hull$ under scaling suggests choosing, within a scaling class, a representative with minimal hull; quantifying the resulting gain for the support-splitting algorithm \cite{S2000,BOS2019} is a natural question.
\end{enumerate}

\section*{Acknowledgments}
This work was carried out during the third author's visit to LAGA (University
Paris~8), funded by the 2025 IMU-Simons Research Fellowship Program for
Developing Countries, within a joint research project proposed by the second
author in the framework of a French--Brazilian scientific cooperation. The
third author was also partially supported by the Fundação de Amparo à Pesquisa
do Estado do Rio de Janeiro (FAPERJ), Brazil, grant CNE E-26/204.037/2024. The
first and third authors were partially supported by the Coordenação de
Aperfeiçoamento de Pessoal de Nível Superior -- Brasil (CAPES), Finance Code 001.

\end{document}